\documentclass{emulateapj}

\submitted{Draft version  \today}
\journalinfo{\@Draft}

\usepackage{natbib}
\usepackage{graphicx}

\usepackage{xcolor}
\usepackage{ amssymb }
\usepackage{color}
\usepackage{comment}
\usepackage{amsmath}
\definecolor{color}{RGB}{25,25,112}
\definecolor{negro}{RGB}{0,0,0}
\definecolor{colorurl}{RGB}{25,25,112}

\usepackage[hyperfootnotes=false]{hyperref}
\hypersetup{
  colorlinks,
  citecolor=color,
  linkcolor=color,
  urlcolor=colorurl}
\shorttitle{Reverse Shock in Candidate SGRB 180418A}

\begin{document}

\title{Reverse Shock Emission Revealed in Early Photometry in the Candidate Short GRB 180418A}
 \author{%
Becerra,~R.~L.$^{1}$;
Dichiara,~S.$^{2,3}$;
Watson,~A.~M.$^{1}$;
Troja,~E.$^{2,3}$;
Fraija,~N.$^{1}$;
Klotz,~A.$^{4}$;
Butler,~N.~R.$^{5}$;
Lee,~W.~H.$^{1}$;
Veres,~P.$^{6}$;
Turpin,~D.$^{7}$
Bloom,~J.~S.$^{2}$; 
Boer,~M.$^{8}$;
Gonz\'alez,~J.~J.$^{1}$;  
Kutyrev,~A.~S.$^{3}$; 
Prochaska,~J.~X.$^{9}$; 
Ramirez-Ruiz,~E.$^{9}$; 
Richer,~M.~G.$^{10}$;  
}

\address{
$^1$ Instituto de Astronom{\'\i}a, Universidad Nacional Aut\'onoma de M\'exico, Apartado Postal 70-264, 04510 M\'exico, CDMX, M\'exico;\\
$^2$Department of Astronomy, University of Maryland, College Park, MD 20742-4111, USA;\\
$^3$Astrophysics Science Division, NASA Goddard Space Flight Center, 8800 Greenbelt Road, Greenbelt, MD 20771, USA;\\
$^4$ Institut de Recherche en Astronophysique et Plan{\'e}tologie, 14 Avenue Edouard Belin, 31400 Toulouse, France;\\
$^5$ School of Earth and Space Exploration, Arizona State University, Tempe, AZ 85287, USA;\\
$^6$ Center for Space Plasma and Aeronomic Research (CSPAR), University of Alabama in Huntsville, Huntsville, AL 35899, USA;\\
$^{7}$ National Astronomical Observatories/Chinese Academy of Science 20A Datun Road, Beijing, 100012, China;\\
$^8$  ARTEMIS, UMR 7250 (CNRS/OCA/UNS), boulevard de l'Observatoire, BP 4229, F 06304 Nice Cedex, France;\\
$^9$Department of Astronomy and Astrophysics, UCO/Lick Observatory, University of California, 1156 High Street, Santa Cruz, CA 95064, USA;\\
$^{10}$  Instituto de Astronom{\'\i}a, Universidad Nacional Aut\'onoma de M\'exico, Unidad Acad\'emica en Ensenada, 22860 Ensenada, BC, Mexico;\\}

\begin{abstract}
We present observations of the possible short GRB 180418A in $\gamma$-rays, X-rays, and in the optical. Early optical photometry with the  TAROT and RATIR instruments show a bright peak ($\approx$ 14.2 AB mag) between $T+28$ and $T+90$ seconds that we interpret as the signature of a reverse shock. Later observations can be modeled by a standard forward shock model and show no evidence of a jet break, allowing us to constrain the jet collimation to $\theta_j> 7^\circ$.
Using deep late-time optical observations we place an upper limit of $r>24$ AB mag on any underlying host galaxy. The detection of the afterglow in the \textit{Swift} UV filters constrains the GRB redshift to $z<1.3$ and places an upper bound on the $\gamma$-ray isotropic equivalent energy $E_{\rm{\gamma,iso}} < 3 \times 10^{51}$ erg.

The properties of this GRB (e.g. duration, hardness ratio, energetic, and environment) lie at the intersection between short and long bursts, and we can not conclusively identify its type. We estimate that the probability that it is drawn from the population of short GRBs is 10\%-30\%.
\end{abstract}
\begin{center}
\keywords{(stars) gamma-ray burst: individual (\objectname{GRB 180418A}).}
\end{center}

\section{Introduction}
\label{sec:introduction}
Gamma-ray bursts (GRBs) are the brightest events in the universe. There are two main populations of GRBs: short GRBs (SGRBs) and long GRBs (LGRBs). The populations are distinguished by their duration $T_{90}$, the interval in the observer's frame over which 90\% of the total background-subtracted counts are observed \citep{Kouveliotou93}, and other secondary parameters such as hardness and spectral lag \citep{2015PhR...561....1K}. The population of SGRBs typically has $T_{90} < 2$ seconds and harder spectra whereas the population of LGRBs typically has $T_{90} > 2$ seconds and softer spectra \citep{2013FrPhy...8..661G}. Nevertheless, there is some overlap between the distributions of $T_{90}$ and hardness, and in some cases it is not clear whether a burst with intermediate properties belongs to the population of SGRBs or LGRBs.

SGRBs are thought to be the consequence of mergers between compact objects driven by angular momentum and energy losses to gravitational radiation \citep[e.g.][]{1989Natur.340..126E,1992ApJ...395L..83N,1998A&A...338..535R,2002MNRAS.336L...7R,2011PhRvD..83d4014G,2007NJPh....9...17L}. The resulting system can result in a stable neutron star, a black hole, or a supra-massive, rotationally supported neutron star which collapses to a black hole due to the lost of angular momentum \citep{2017ApJ...844L..19P}. The gravitational waves emitted during the merger provide stringent constraints on the individual masses of the coalescing objects \citep{2017PhRvL.119p1101A}. The first discovery of a binary neutron star merger (GW170817) in gravitational waves revealed the importance of the multi-messenger approach. Combined detection of electromagnetic counterparts (the kilonova, GRB prompt and the afterglow) is crucial for the understanding of these peculiar phenomena \citep[][]{2017ApJ...848L..12A}.

On the other hand, most LGRBs arise from the core collapse of massive stars \citep{2003Natur.423..847H} and are associated with hydrogen-poor, high-velocity type Ic supernovae \citep{2013MNRAS.434.1098C}. 

According to the standard fireball model \citep{1993ApJ...418L...5P,1999PhR...314..575P,2004IJMPA..19.2385Z,2015PhR...561....1K} in both SGRBs and LRGBs, the electromagnetic emission is produced by collimated relativistic ejecta, which initially interacts with itself and then later with the circumburst medium

Two distinct emission phases occur in the GRB fireball scenario. First, there is a prompt phase producing the short-lived gamma-ray radiation through internal shocks within the relativistic jet that dissipate its internal kinetic energy. Then, there is an afterglow phase during which a longer fading multi-wavelength emission is radiated from the external shocks between the jet and the circumstellar medium \citep{2015PhR...561....1K,1999PhR...314..575P}. Two kinds of external shocks are important: long-duration forward shocks which propagate outward sweeping up the circumstellar medium and a short-lived reverse shocks which propagate backward into the jet \citep{1993ApJ...405..278M}.

Emission from forward shocks explains the afterglow phase of many GRBs. The dynamics of the forward shocks have been amply explored \citep{1997ApJ...476..232M,1999ApJ...520..641S,2002ApJ...568..820G}. Detailed studies of the afterglow emission and especially the forward shock component, provide valuable information about the total energy, geometry, and the structure of the circumburst medium \citep[e.g.][]{2007A&A...474..827S, 2010ApJ...709L.146D, 2006ApJ...653..468B, 2016ApJ...827..102T, 2019ApJ...871..200F}. 
On the other hand, the reverse shock emission is useful for understanding the initial bulk Lorentz factor, the ejecta composition and magnetization \citep[e.g.][]{2009Natur.462..767S, 2013Natur.504..119M, 2016ApJ...818..190F}.

Reverse shocks are discussed by \cite{1997ApJ...476..232M,1999ApJ...520..641S,2000ApJ...545..807K,2015AdAst2015E..13G} and are predicted to generate a strong optical flash observable in the very early stages of the afterglow due to synchrotron emission which is shown as a single-peak \citep{2018ApJ...859...70F} for a moderately magnetized ejecta \citep{2015ApJ...798....3Z}. After the flare no new  electrons  are  injected and the shell material cools adiabatically. 

Rapid (within minutes) and sensitive optical observations are crucial to study the reverse shock emission. Bright optical flashes have been observed in many LGRBs \citep[e.g.][]{1999Natur.398..400A, 2009ApJ...691..723B, 2014Sci...343...38V, 2017Natur.547..425T, 2018ApJ...859...70F} and are inconsistent with a simple forward shock scenario. They are normally explained as emissions emerging from a reverse shock \citep{2014ApJ...785...84J, 2017ApJ...848...94F}. Reverse shock components in the afterglows of SGRBs have been previously suggested on the basis of gamma and radio observations \citep[e.g.][]{2006ApJ...653..468B, 2018Galax...6..103L,2019ApJ...871..123F, 2016ApJ...831...22F,2019arXiv190501290T}, but they have not previously been identified in the optical. Early radio flares have been attributed to the emission from a short-lived external reverse shock \citep{1999ApJ...520..641S,2000ApJ...542..819K}. The particles that produce the optical flash cool adiabatically, and therefore their emission shifts to lower frequencies, over time \citep{2000ApJ...542..819K, 2019arXiv190501290T}. 

In this work, we present the photometric data and analysis of GRB 180418A, detected with TAROT 28 seconds after the gamma-ray trigger. We show that it is the first possible SGRB showing reverse shock emission in the optical. We are unable to rule out the possibility that it is a LGRB.

The paper is organized as follows. In \S\ref{sec:observations}, we present the observations with {\itshape Swift}, {\itshape Fermi}, TAROT, RATIR and other telescopes. In \S\ref{sec:analysis} we present a temporal and spectral analysis and discuss the nature of this GRB. In \S\ref{sec:discussion} we summarize and discuss our results. In Appendix~\ref{app:host}, we present our search for the GRB host galaxy.

\section{Observations}
\label{sec:observations}

\subsection{Neil Gehrels Swift Observatory}
\label{sec:swift}

\begin{figure}
\centering
\includegraphics[width=0.45\textwidth]{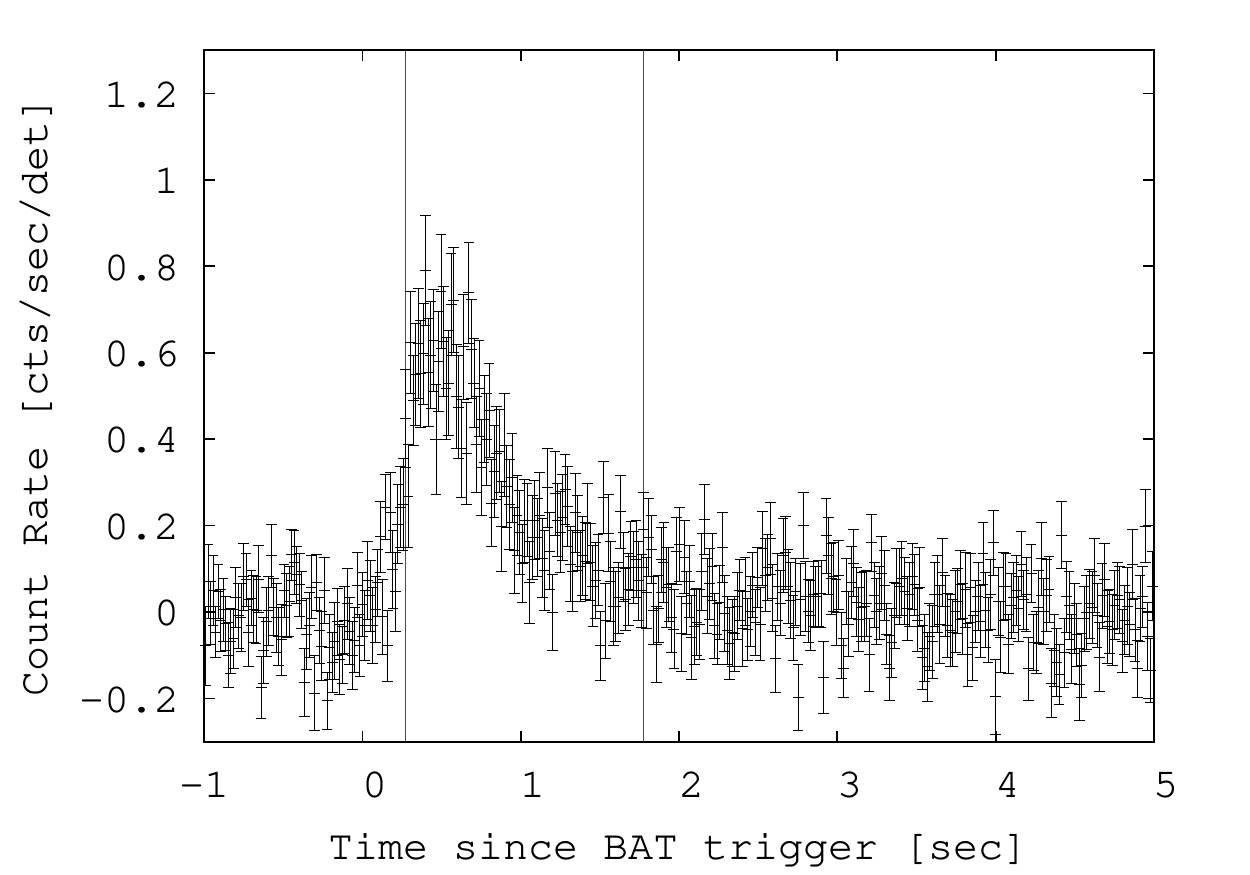}
\caption{{\itshape Swift}/BAT light curve using a binning of 16 ms and observed in the range 15-350 keV. The red lines mark the start and end of the $T_{90}$ interval.} 
\label{fig:swift}
\end{figure}

The {\itshape Swift}/BAT instrument triggered on GRB 180418A at $T=$ 2018 March 25 06:44:06.012 UTC (trigger 826428) \citep{22646}. The BAT light curve showed a single FRED-like pulse that started
at $T+0$ seconds, peaked at about $T+0.4$ seconds, and ended at about $T+3.5$ seconds. The BAT light curve is shown in Figure~\ref{fig:swift}.

The {\itshape Swift}/BAT data of GRB 180418A were processed using HEASOFT package (v6.25). The energy calibration was applied with BATECONVERT and the mask weighting was included with BATMASKWTEVT. We used BATTBLOCKS to run the Bayesian Block algorithm over the 16 ms, background-subtracted, 15--350 keV light curve and determined $T_{90} = 1.504 \pm 0.380$. BATTBLOCKS was run with the default configuration options except the background-subtraction parameter \textit{bkgsub} was set to `YES'. 

The 15--150 keV spectrum integrated over the $T_{90}$ interval (from 0.272~s to 1.776~s) 
is well fitted by a simple power law with a photon index of $1.43 \pm 0.11$ ($\chi^{2}/\mathrm{d.o.f.} = 1.02$) and a fluence of $(2.72 \pm 0.11) \times 10^{-7}\ \mathrm{erg\,cm^2}$. From this spectrum we derive a hardness ratio of $S(100-50)/S(25-50) = 1.48$.

The 15--150 keV peak flux was derived from the spectrum integrated in the interval from 0.144~s to 1.144~s. The peak spectrum is well fitted by a power law with a photon index of $1.41 \pm 0.11$, a peak flux of $(2.36 \pm 0.17) \times 10^{-7}\ \mathrm{erg\,cm^{-2}\, s^{-1}}$, and a peak photon flux of $(2.98 \pm 0.17)\ \mathrm{cm^{-2}\,s^{-1}}$.


Due to an observing constraint, {\itshape Swift} did not slew to the source until $T + 49.6$ minutes \citep{22646}, and therefore, the {\itshape Swift}/XRT instrument only started observing the field at $T+3081.4$ seconds. It detected a fading source at RA DEC 11:20:29.17 +24:55:59.1 J2000 with a 90\% uncertainty radius of 1.8 arcsec \citep{22649, 22650, 22655}. The X-ray spectrum could be fitted with an absorbed
power-law with a photon spectral index of $2.02^{+0.28}_{-0.26}$ and an
absorption column of $8.0^{+7.1}_{-5.8} \times 10^{20}~\mathrm{cm^{-2}}$, in
excess of the Galactic value of $1.1 \times 10^{20}~\mathrm{cm^{-2}}$ \citep{22655}. For our analysis in this paper, we used the X-ray light curve and the spectrum obtained from the pipeline of \cite{2007AJ....133.1027B} and \cite{2007ApJ...671..656B}.

The {\itshape Swift}/UVOT instrument started observing the field at $T+3086$ seconds and detected a fading source at RA DEC
11:20:29.21 +24:55:59.2 J2000 with a 90\% uncertainty radius of 0.49 arcsec \citep{22665}. We downloaded the UVOT data from the online archive \footnote{https://swift.gsfc.nasa.gov/archive/} and derived magnitudes and signal-to-noise ratios.
Table~\ref{tab:datosuvot} shows the filter, initial time $t_\mathrm{i}$ and the final time $t_\mathrm{f}$ (relative to $T$), the AB magnitude, and the signal-to-noise ratio. For UVOT, the exposure time is simply $t_\mathrm{f}- t_\mathrm{i}$.


\subsection{Fermi Gamma-Ray Observatory}
\label{sec:fermi}

\begin{figure}
\centering
\includegraphics[width=0.45\textwidth]{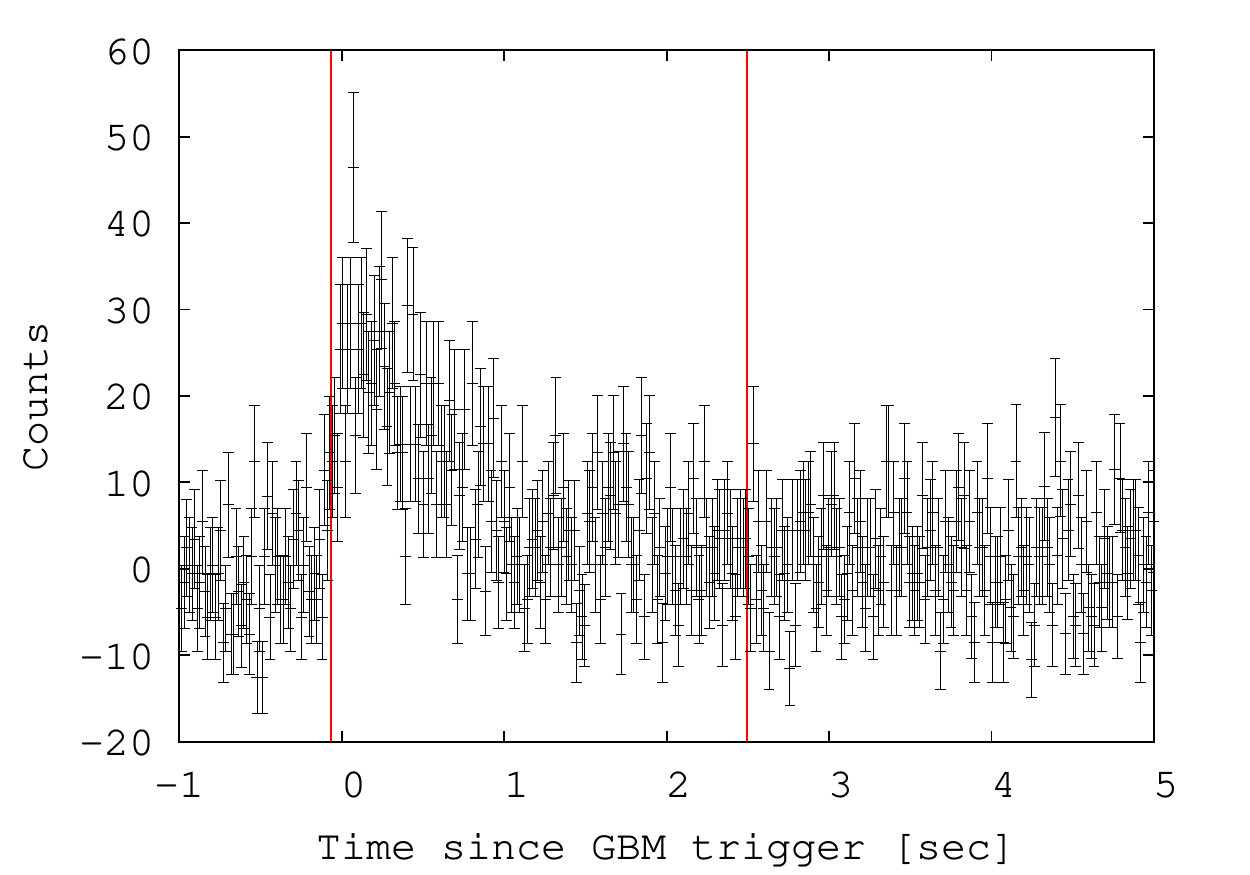}
\caption{GBM light curve since the GBM trigger. The energy range is 8-1000 keV and the binning of 16 ms. The red lines mark the start and end of the $T_{90}$ interval.}
\label{fig:gbm}
\end{figure}

The {\itshape Fermi}/GBM instrument triggered on GRB 180418A at 2018 April 18 06:44:06.28 UTC (trigger 545726651/180418281) and observed a single FRED-like peak (Figure \ref{fig:gbm}), in agreement with the {\itshape Swift}/BAT light curve, with a $T_{90}$ duration of $2.56 \pm0.20$ seconds and a 10-1000 keV fluence of $(5.9\pm0.1) \times 10^{-7}~\mathrm{erg\,cm^{-2}}$ \citep{22656,2016ApJS..223...28N}. The GBM light curved is shown in Figure~\ref{fig:gbm}. The burst was not detected by the {\itshape Fermi}/LAT instrument.




\subsection{TAROT Observations}
\label{sec:observationstarot}

TAROT\footnote{\url{http://tarot.obs-hp.fr/}} La Silla is a 25-cm robotic telescope located at the European Southern Observatory,
La Silla Observatory, in Chile. TAROT is equipped with a CCD camera and a filter wheel with $BVRIC$ filters \citep{2008PASP..120.1298K}.

TAROT is connected to the GCN/TAN alert system and received a BAT quick look alert packet for GRB 180418A at 06:44:21 UTC ($T + 15$ seconds). It immediately slewed to the burst and began observing in the clear $C$ filter, with the first exposure starting at 06:44:34 UTC ($T+28$ seconds). The first exposure is trailed with a duration of 60 seconds to allow continuous monitoring of the light curve \citep{2006A&A...451L..39K}. Subsequent exposure were taken in sidereal tracking mode with exposure times of 30 to 90 seconds and read-out times of about 10 seconds. We use TAROT data from $T+28$ seconds to $T+392$ seconds \citep{22671}.



Table~\ref{tab:datostarot} gives TAROT photometry. For each exposure, it gives the initial time $t_\mathrm{i}$ and the final time $t_\mathrm{f}$ (relative to $T$) and the AB $r$ magnitude (obtained from the $C$ magnitude and not corrected for Galactic extinction) with the the 1$\sigma$ total uncertainties (including both statistical and systematic contributions). For TAROT, the exposure time is simply $t_\mathrm{f}-t_\mathrm{i}$. 


\subsection{RATIR Observations}
\label{sec:observationsratir}

RATIR\footnote{\url{http://ratir.astroscu.unam.mx/}} is a four-channel simultaneous optical and near-infrared imager mounted on the 1.5-meter Harold L.\ Johnson Telescope at the Observatorio Astron\'omico Nacional in Sierra San Pedro M\'artir in Baja California, Mexico \citep{2012SPIE.8446E..10B,2012SPIE.8444E..5LW,2015MNRAS.449.2919L}. RATIR usually obtains simultaneous photometry in $riZJ$ or $riYH$, but at the time of these observations the $ZY$ and $JH$ channels were not operational. Therefore, RATIR only obtained photometry in $ri$.

RATIR is connected to the GCN/TAN alert system and received a BAT quick look alert packet for GRB 180418A at 06:44:20.3 UTC ($T + 14.3$ seconds). It immediately slewed to the burst and began observing, with the first exposure starting at 06:46:06.3 UTC ($T+120.6$ seconds). It took simultaneous exposures in $r$ and $i$ with an exposure time of 80 seconds and a cadence of about 100 seconds. On the nights of 2018 April 18, 19, 20, and 21 UTC, RATIR observed from $T + 120.6$ seconds to $T + 3.64$ hours \citep{22652}, from $T+ 20.47$ to $T + 27.09$ hours \citep{22664}, from $T + 44.62$ to $T + 51.37$ hours, and from $T+68.55$ to $T+ 75.27$ hours.

The RATIR reduction pipeline performs bias subtraction and flat-field correction, followed by astrometric calibration using the \url{astrometry.net} software \citep{2010AJ....139.1782L}, iterative sky-subtraction, coaddition using SWARP, and source detection using SEXTRACTOR \citep{2015MNRAS.449.2919L}. Images were calibrated against SDSS \citep{2015MNRAS.449.2919L}.

Table~\ref{tab:datosratir} gives our RATIR photometry. For each exposure or coadded exposure it gives the initial time $t_\mathrm{i}$ and final time $t_\mathrm{f}$ (relative to $T$), the total exposure time $t_\mathrm{exp}$, and the $r$ and $i$ magnitudes (not corrected for Galactic extinction) with their 1$\sigma$ total uncertainties (including both statistical and systematic contributions). 

\begin{figure*}
\centering
 \includegraphics[width=0.8\textwidth]{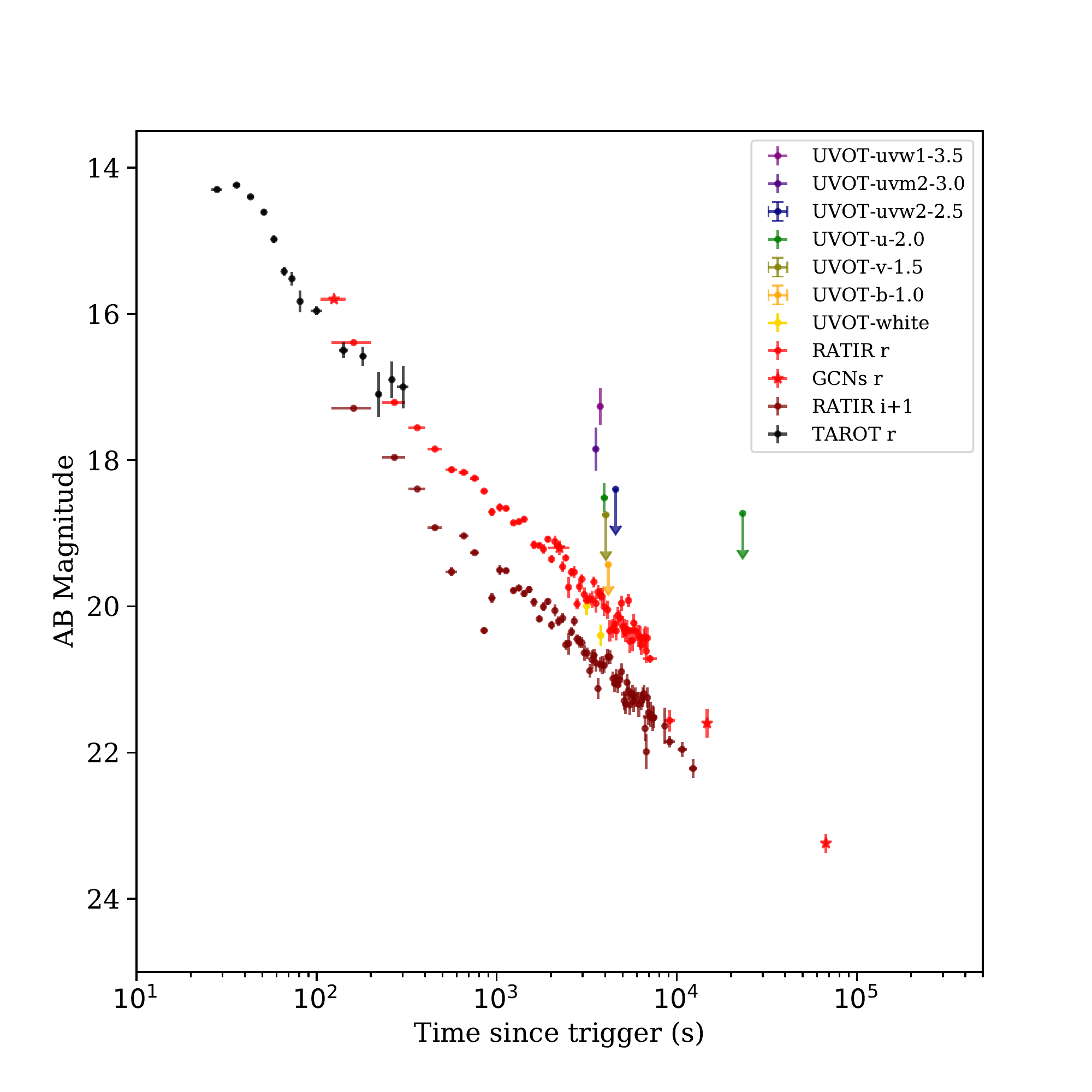}
 \caption{GRB 180418A photometry from this work (TAROT, RATIR, and UVOT) and GCNs \citep{22647,22648,22659,22662}}
 \label{fig:observations}
\end{figure*}

\subsection{Other Terrestrial Observations}
\label{sec:observationsterrestrial}

\cite{22647} began to observe the field of GRB 180418A with KAIT at 06:46:41 UTC ($T+ 125$ seconds). They reported an optical transient that faded from about magnitude 15.8 at $T + 135$ seconds. Due to the delay in the {\itshape Swift}/XRT observations, this was the first precise localization of the counterpart of GRB 180418A. Additional photometry was reported by \cite{22648,22657,22659,22660,22662,22663,22666,22668,22670}

\cite{22662} mentioned that the source was not clearly point-like in their images and suggested this might be due to the presence of a host galaxy close to the afterglow. However, our RATIR observations from 2018 April 19 ($T+68.5$ to $T+75.3$ hours) revealed no detection to a $3\sigma$ limiting magnitudes of $r > 24.0$ and $i> 23.9$, which places a limit on the magnitude of any close host galaxy. We discuss in details our analysis of the host galaxy search and upper limits in \S~\ref{app:host}.

Our TAROT and RATIR observations combined with other public data are shown in Figure~\ref{fig:observations}.


\section{Analysis}
\label{sec:analysis}

\subsection{Temporal Analysis}

The prompt emission from the GRB detected by {\itshape Fermi}/GBM and {\itshape Swift}/BAT lasted until about $T + 2.5$ seconds. The earliest data from {\itshape Swift}/XRT start at $T + 52$ minutes. Our optical observations from TAROT and RATIR begin at $T + 28$ and $T + 121$ seconds, respectively, significantly after the end of prompt emission. Therefore, we focus our analysis on the afterglow.

\begin{figure*}
\centering
 \includegraphics[width=0.45\textwidth]{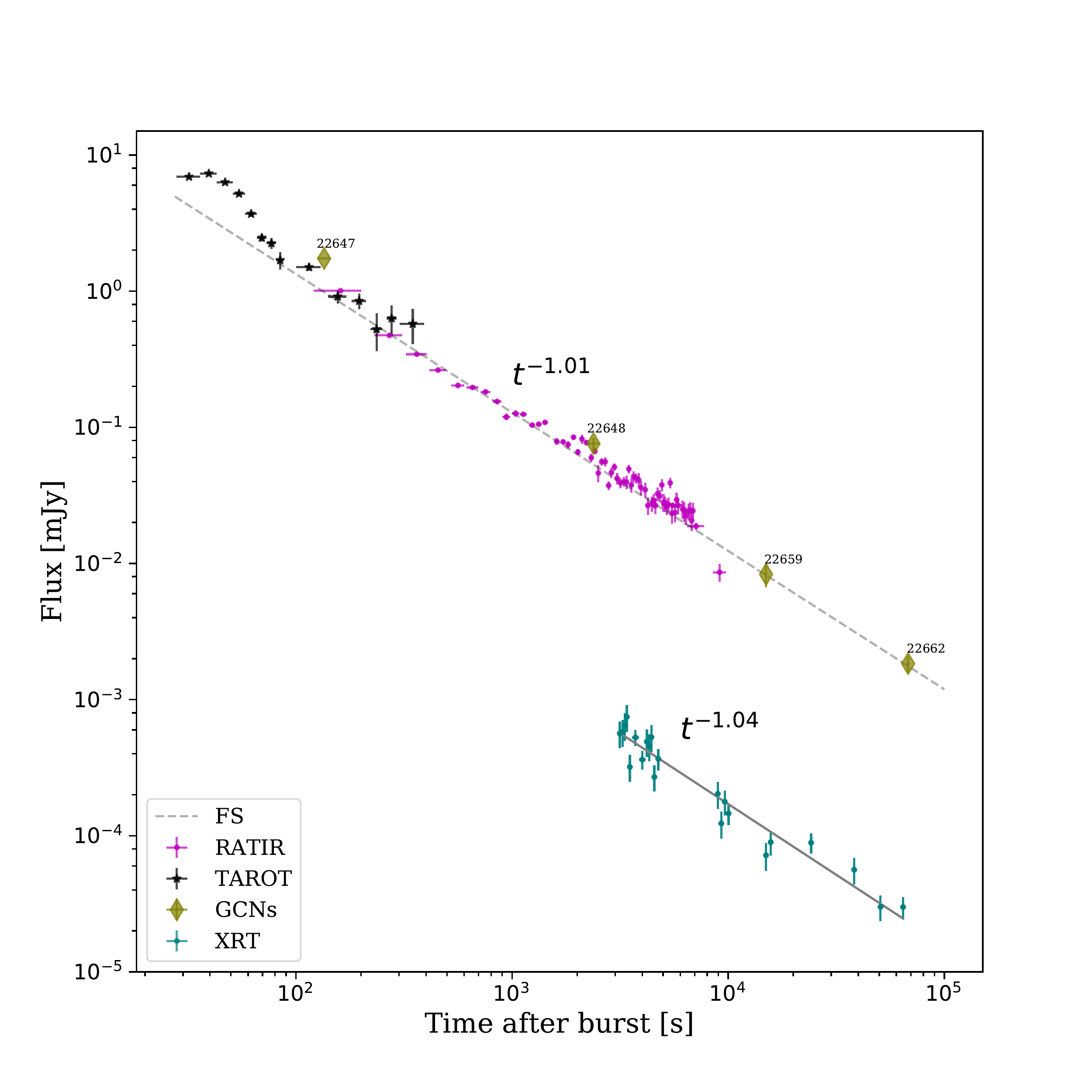}
 \includegraphics[width=0.45\textwidth]{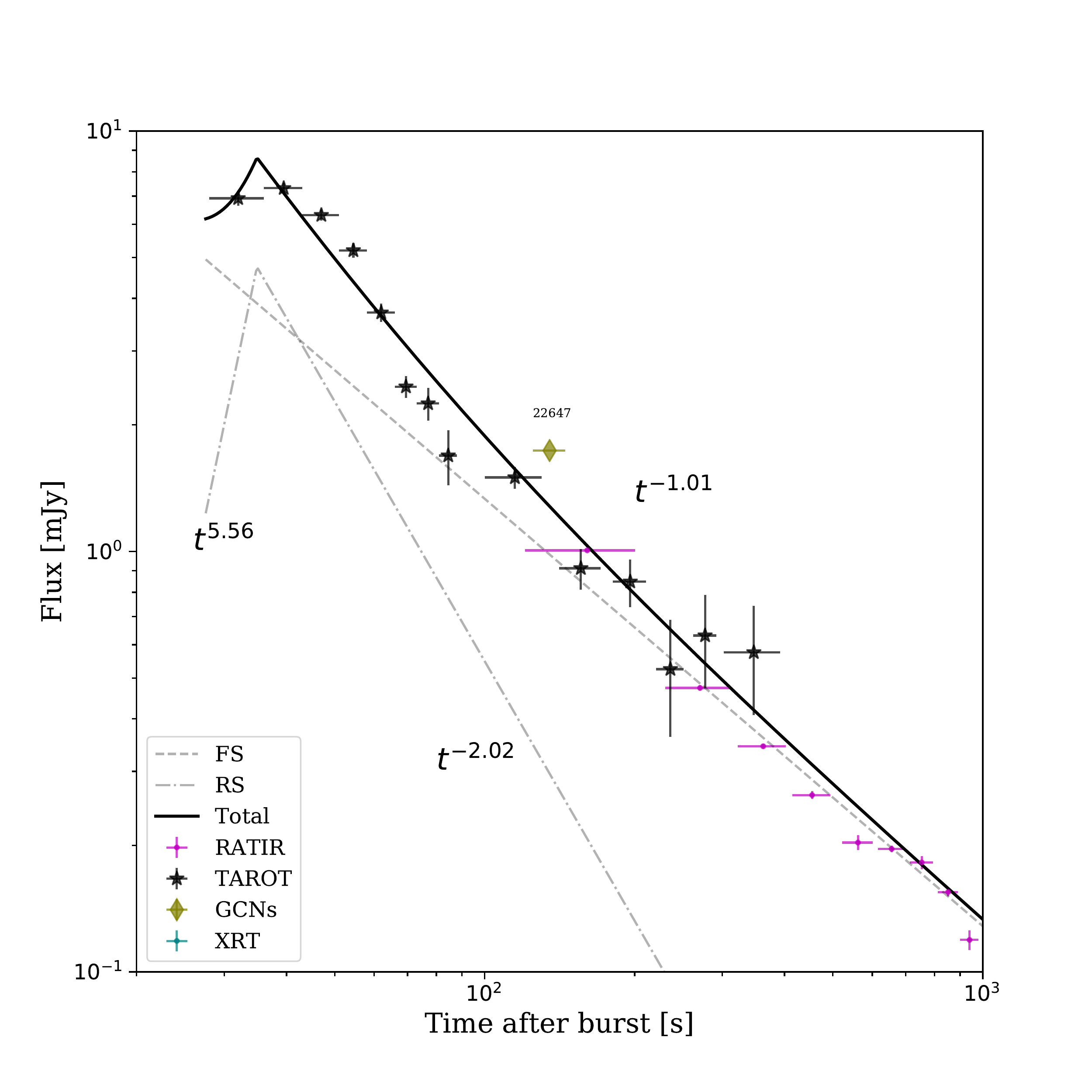}
 \caption{Left: Fitting of the GRB 180418A light curve with a single power-law. Points red and blue are observations in $r$-band and $X$-rays respectively. The X-ray and r-band show a non break up to $10^5$s. Right: Zoom of GRB 180418A light curve at early times with a fit using a reverse (RS) and a forward shock (FS) components The final fit for optical is the black line (which includes the sum of forward and reverse shock)}
 \label{fig:lightcurve}
\end{figure*}

\subsubsection{Optical Temporal Analysis}
\label{sec:optical-temporal-analysis}

We use TAROT/RATIR $r$-filter data for the analysis, as we have more data in this band and the $i$-filter of RATIR shows the same qualitative behavior. We complement our data set with information published in GCNs \cite{22647,22648,22659,22662}. 
Figure~\ref{fig:lightcurve} shows the optical and X-ray light curves for GRB 180418A.




The optical light curve appears to be a smooth power-law decline but with an excess of emission at early times up to about 100 seconds.
The later standard forward shock decay is present in most afterglows \citep{2006ApJ...642..354Z}. 

We fit the optical light curve for $t > 1000$ seconds using a simple forward external shock component into a constant-density ISM under the supposition of a thin-shell evolving in the slow-cooling regime below the cooling break \citep{2000ApJ...545..807K}, and find a temporal index of $\alpha_\mathrm{forward} = 1.01\pm0.01$. The temporal index $\alpha_\mathrm{forward}$ is expected to be related to the electron energy index $p$ by $\alpha_\mathrm{forward} =3(p-1)/4$, so we predict $p = 2.35 \pm 0.01$. A jet being driven into a decreasing-density stellar-wind environment \citep{2017ApJ...848...15F} would give a $p < 2$, which is not a typical value observed for GRBs and then, we discard this possibility for this work although we cannot reject that unlikely scenario totally.

The light curve for the forward-shock component is shown in the left panel of Figure~\ref{fig:lightcurve}. However, extrapolating this fit to earlier times, is it clear that there is a significant excess of emission, and this is shown in the right panel of Figure~\ref{fig:lightcurve}. This excess is present from our earliest observation at $t=28$ seconds until $t\approx100$--300 seconds and has a peak at $t\approx40$ seconds. A priori, there are several different scenarios which might explain this excess. 

First, we consider the possibility of late central engine activity. This scenario has been studied widely for X-rays.  Figures~\ref{fig:lightcurve} and \ref{fig:sed} suggest an achromatic behavior for the X-ray and optical emission and, therefore, the same region of origin.
In this case, we would expect a counterpart in X-ray for our detected excess. 
Unfortunately, we do not have data from XRT for these times.

\cite{2017ApJ...844...79Y} suggest that a possible difference between the duration of X-ray and UV/optical flares is because we only see the peak of the flare in the UV/optical and therefore, our measured duration for the flare will be biased relative to the X-ray where we see more of the flare rise and decay but not their temporary indices \citep{2014ApJ...788...30S}.  For the optical, the timing of the excess corresponds to a range of the relative duration $\delta t/t$ (duration of about 100 seconds over the peak time of about 40 seconds) of more than 2. Therefore, we would expect a $\delta t/t\ge2$ for the predicted X-ray counterpart and this is gone against the typical yields for X-rays calculated for late central activity of $\delta t/t\ll 1$ \citep{2006ApJ...642..354Z}. 
 In conclusion, the late central activity scenario is unlikely to power the observed rebrightening observed in GRB 180418A 
 
Second, we consider a two-component jet scenario consisting of a narrow and initially faster component and a wide and initially slower component. This model was proposed to explain the rebrightening emission in afterglow light curves \citep{2005ApJ...631.1022G, 2008Natur.455..183R}. \cite{2005ApJ...631.1022G} derived the two-component jet light curves and found that this model cannot produce very sharp features in the light curve. Also they concluded that this scenario typically show a pronounced brightening detected in light curves of several afterglows 0.1-1 days after the GRB (depending on the environment) according to analytic solutions for typical microphysical parameters found in most GRBs (p=2.25, $\epsilon_e=0.01$,, $\epsilon_B=0.001$, $n=0.2$ cm$^{-3}$) \citep{2005ApJ...626..966P}.
Nevertheless, in that case, the excess would appear at least couple of hours after the trigger, and so this cannot explain a very early excess like the one seen here in GRB 180418A. 

Finally, we consider emission from a reverse external shock \citep{2000ApJ...545..807K} in addition to a standard forward shock. The optical light curve of GRB 180418A is similar to the a \emph{Peak + Fast Decay} curves studied by \cite{2010ApJ...720.1513K} which show a fast rise to a peak, followed by a fast decay (with a temporal index $\alpha$ of about 1.5-2) and typically interpreted to a reverse shock component extra to the forward component \citep{2014ApJ...785...84J} or in some cases as central engine late activity.
For this model, the temporal indices of the rise and the decay are related to the peak time of the emission of the reverse shock $t_\gamma$ and the electron index $p$ (assumed to be the same for both the forward and reverse shocks) by $\alpha_\mathrm{reverse,R} = 3/2-3p$ and $\alpha_\mathrm{reverse,D} = (27p+7)/35$, respectively. Thus, the free parameters of the model for the total emission from both shocks are $p$, $t_\gamma$, and the normalization of the two components.  The final fit for both components is shown in the right panel of Figure \ref{fig:lightcurve} and has a $\chi^2/d.o.f.=0.98$ with an electron index $p=2.35\pm0.03$. The parameters for this model are shown in Table~\ref{tab:fit}. (The temporal indices $\alpha$ are labeled with suffixes for reverse and forward and with $R$ or $D$ to refer to the rise and decay phase of the reverse component.)



\subsubsection{X-Ray Temporal Analysis}

{\itshape Swift/}XRT data were obtained for $t > 51$ minutes. This region can be fitted as a power-law with a temporal index of $\alpha_\mathrm{X,forward} = 1.04 \pm 0.05$. The similarity of the X-ray and optical temporal indices suggests that the emission in both wavelength regions arises from the same spectral regime, which we assume to be a thin-shell evolving in the slow-cooling regime below the cooling break into a constant density ISM \citep{2000ApJ...545..807K}.

\subsection{Spectral Analysis}

\begin{figure}
\centering
 \includegraphics[width=0.5\textwidth]{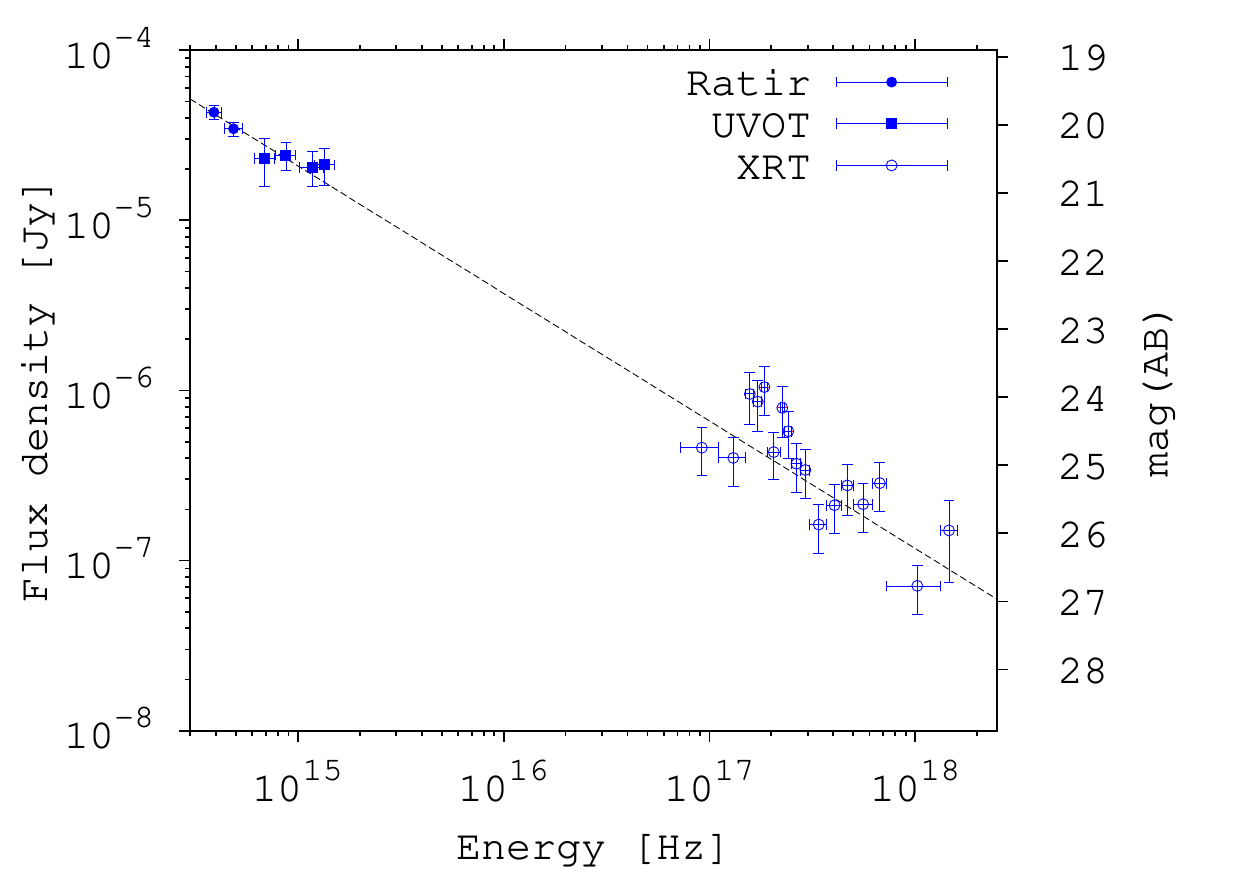}
 \caption{The SED of GRB 180418A from X-rays to the optical at $t = 4000$ seconds. The data are from {\itshape Swift/}XRT, {\itshape Swift/}UVOT, and RATIR. The dotted line is a simple power law with index $\beta=0.73$. The solid line is the best fit from XSPEC.}
 \label{fig:sed}
\end{figure}


We retrieved the {\itshape Swift/}XRT X-ray spectrum in the time interval 3091--4831 s
from the online repository \footnote{http://www.swift.ac.uk/xrt\_spectra/}.
We interpolated our UVOT and RATIR photometry to $t=4000$ s assuming a power-law decay with an index of $\alpha_\mathrm{forward} = 1.01$ (see \S\ref{sec:optical-temporal-analysis}). Figure~\ref{fig:sed} shows the resulting combined spectral energy distribution (SED).

From the X-ray to the optical, the SED can be fitted with a simple power law with index $\beta=0.73\pm0.03$. Under our assumption of a thin-shell evolving in the slow-cooling regime with the cooling break above the X-rays \citep{2000ApJ...545..807K}, we would expect the spectral index to be $(p-1)/2$ or $0.68\pm0.03$ for $p=2.35$. Thus, our observed value is in good agreement with this prediction. 

\subsection{Photometric Redshift}
\label{sec:redshift}

No spectroscopic redshift for the burst has been reported, therefore we are forced to place limits on the redshift from the combined X-ray, UV, and optical broad-band SED. The  SED  fitting  was  carried  out  using XSPECv12.10.1 \citep{1996ASPC..101...17A} taking into account Galactic extinction for IR, optical and UV energy bands due to dust \citep{1989ApJ...345..245C}, the IR/optical/UV extinction of the host galaxy, and the photoelectric absorption of soft X-rays (in the Galaxy and in the host galaxy). We set the redshift as a free parameter during the fit, and found an upper limit of $z < 1.31$ at a 90\%  confidence level. This limit basically comes from the fit requiring that the Lyman continuum absorption fall below our detection in the UVOT $uvm2$ filter.

\subsection{Classification of the Burst}
\label{sec:classification}

The values of $T_{90}$ do not unambiguously identify GRB 180418A as a SGRB or LGRB because is not a hard neither definitive criteria to distinguish between the possible progenitors. The value of $T_{90}=1.50 \pm 0.38$ from the {\itshape Swift}/BAT light curve would suggest a SGRB, although it is only below 2 seconds by $1.3\sigma$. However, the value of $T_{90}=2.56\pm0.20$~s from the {\itshape Fermi}/GBM light curve \citep{22656} would suggest a LGRB. 

Additional information on the nature of a GRB can sometimes be obtained from the spectral hardness and lag. The spectral hardness found from {\itshape Swift}/BAT instrument is $S(100-50)/S(25-50)=1.48$, which corresponds to intermediate hardness in the classification of \cite{2013ApJ...764..179B} and does not help to resolve the ambiguity. Considering both $T_{90}$ and the spectral hardness with the distributions presented by \cite{2013ApJ...764..179B}, we estimate the probability that the burst is a SGRB is 10\%--30\%.

\begin{figure}
\centering
 \includegraphics[width=0.45\textwidth]{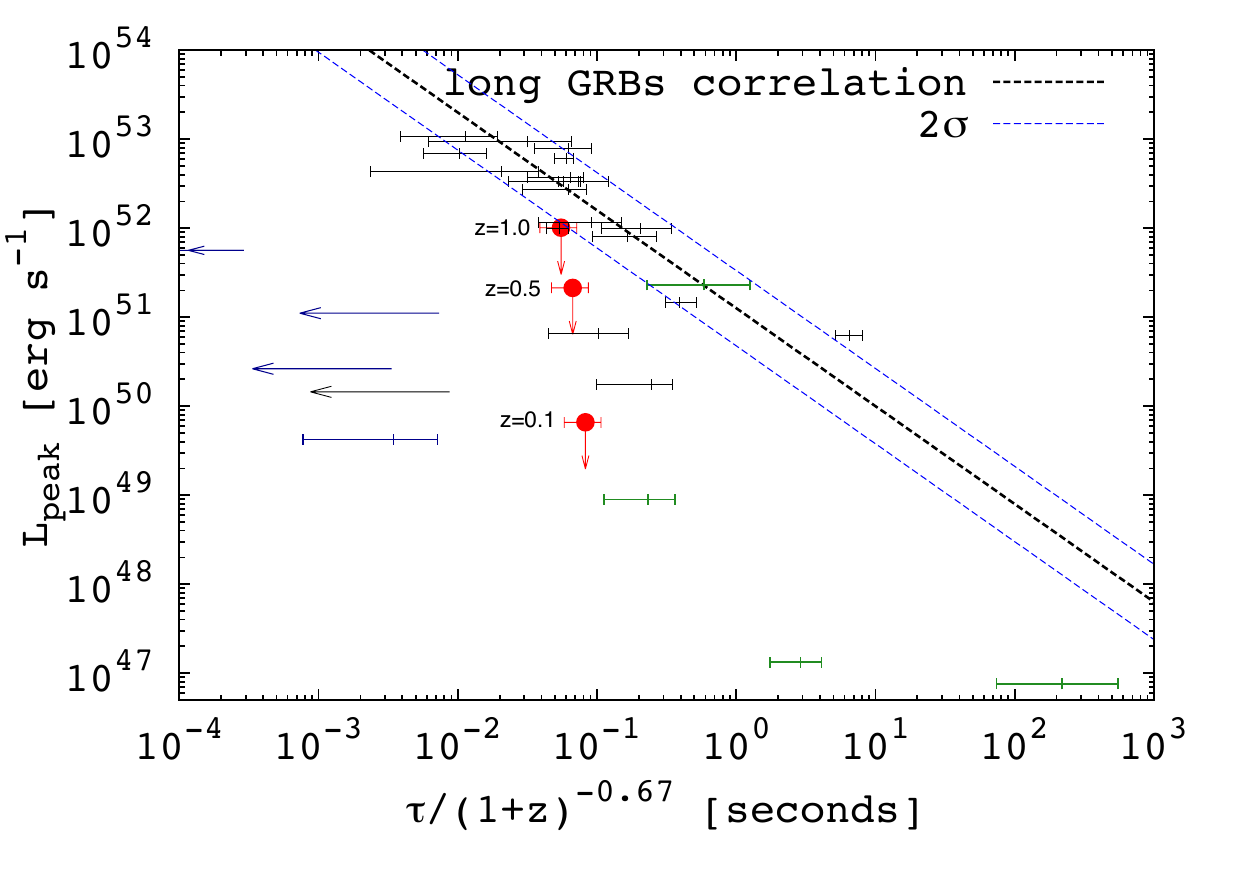}
 \caption{Relation between $\tau$ and $L_{peak}$. The red data show the values for  GRB 180418A assuming redshifts of z=0.1, z=0.5, z=1.0. The black dotted line is the relation from \cite{2000ApJ...534..248N} and the blue dotted lines show the 2$\sigma$ scatter. The figure also shows SGRBs from \cite{2006Natur.444.1044G} including the energy band correction.}
 \label{fig:lag}
\end{figure}

We also investigate the lag between the arrival times between high- and low-energy photons $t_{lag}$ and peak luminosity $L_{peak}$ correlation for LGRBs reported by \cite{2000ApJ...534..248N}.
We used RMfit\footnote{https://fermi.gsfc.nasa.gov/ssc/data/analysis/rmfit/}, version 432, to analyze the spectrum around the peak of the GBM light curve in the time interval from $-0.128$ to 0.896 s. The spectrum fits with a single power-law with index $-1.46\pm0.07$ and flux = $(6.0 \pm 1.2) \times 10^{-7}$ between 10 and 1000 keV. Using these parameters we derived an upper limits for the peak luminosity $L_{peak}$ as a function of redshift, obtaining $L_{peak} = 6.57 \times 10^{49}$~erg s$^{-1}$, $L_{peak} = 2.14 \times 10^{51}$~erg s$^{-1}$ and  $L_{peak} = 1.01 \times 10^{52}$~erg s$^{-1}$ for z=0.1, z=0.5 and z=1.0, respectively. The time-lag between the energy bands 50-100 keV and 15-25 keV ($\tau$ = $0.088 \pm 0.026$) was retrieved from the GCN notice 22658 \cite{22658}. We applied the time dilation correction and energy correction due to the redshift considering the relation between the pulse width and energy \citep{2002ApJ...579..386N} and approximating the correction factor with $(1+z)^{0.67}$ as proposed by \cite{2006Natur.444.1044G}. The results of this analysis are shown in Figure~\ref{fig:lag} and compared with data of SGRBs presented by \cite{2006Natur.444.1044G} and the LGRB correlation of \cite{2000ApJ...534..248N}. We see that GRB 180418A is consistent with an LGRB provided the redshift is larger than about 1 and is also consistent with a SGRB at lower redshifts. Unfortunately, our photometric analysis in \S\ref{sec:redshift} only requires $z < 1.3$, and so we are unable to rule out the possibility and again cannot conclusively decide on the nature of the burst.

In conclusion, neither $T_{90}$ nor the spectral hardness nor the spectral lag allow us to conclusively determine if GRB 180418A is drawn from the populations of SGRBs or LGRBs. Nevertheless, we constrain the redshift as z$>$0.3 (see Appendix~\ref{app:host}) and z$<$1.3 from the SED. For this specific case, the non-detection of a supernova associated to this event, in distances where telescopes do not have sensibility to detected it and, the no identification of host galaxy cannot support nor deny the idea to eliminate the possibility to have a merger of two compact objects or the death of a massive star as progenitor GRB 180418A.

\subsection{Physical parameters}
\label{sec:parameters}

We can estimate some intrinsic parameters of the burst assuming a $\Lambda$CDM cosmology with a $H_0=67.8$ km/Mpc/s \citep{2014A&A...571A...1P}. 
For a redshift value of $z < 1.3$ we obtained a luminosity distance of $\lesssim $9 Gpc and $E_{\gamma,iso} \lesssim 3\times 10^{51}$erg which is in the typical range for short GRBs \citep{2014ARA&A..52...43B}. The relation of optical luminosity for SGRBs at a rest-frame time of 7h ($L_{opt,7}$) as function of $E_{\gamma,iso}$ according to \cite{2014ARA&A..52...43B} is given by:
\newcommand{\ergs}{\ensuremath{\,\textrm{erg s$^{-1}$}}}

\begin{equation}
L_{opt,7}\approx 6.9 \times 10^{42}E_{\gamma,iso,51}^{0.74}\quad \ergs,
\end{equation}
then $L_{opt,7}\lesssim 10^{43}$erg s$^{-1}$, in agreement with the value reported in Table~\ref{tab:fit}.

The multi-wavelength afterglow is described by a simple power-law decay for $t > 100$ sec, and shows no evidence of a jet break. The relation between the jet break time and half opening angle $\theta_j$ is related with density, energy, time of break and redshift as described in \cite{1999ApJ...519L..17S}:
\begin{equation}
\theta_j=0.13\left (\frac{t_j}{1+z}\right )^{3/8}\left (\frac{n}{E_{52}}\right )^{1/8},
\end{equation}
which, in the case of GRB 180418A, leads to a constraint of
$\theta_j$\textgreater$7^\circ$. Taking a z=0.5 (the typical redshift of SGRBs \cite{2014ARA&A..52...43B}), we calculate also $\theta_j$\textgreater$7^\circ$.

Within the framework discussed by \cite{2017ApJ...848...15F}, the afterglow parameters were calculated using the $\chi^2$ minimization within the ROOT software package \citep{1997NIMPA.389...81B} and following the model from \cite{1999ApJ...517L.109S,1999ApJ...518L..73F,2002MNRAS.330L..24S,2003ApJ...582L..75K,2007ApJ...655..391K}.


All these values are reported in Table~\ref{tab:fit2}.

\section{Discussion and Summary}
\label{sec:discussion}

We have presented a multi-wavelength study of GRB 180418A. Thanks to the rapid response of TAROT, we were able to detect the optical counterpart as early as 28 seconds after the burst. We model the early and late light curves as emission from forward and reverse shocks in the standard fireball scenario. The reverse shock component peaks at $T+ 35$ seconds at an observed magnitude of $r=14.2$.

We calculated the afterglow parameters and constrained the jet collimation to $\theta_j>7$~deg.

Optical emission from reverse shocks has been seen in many LGRBs (see  Table~\ref{tab:list}), but never in SGRBs. Unfortunately, although GRB 180418A is consistent with a SGRB, it is also consistent with an LGRB with $1.0 < z < 1.3$. The absence of a spectroscopic redshift makes it impossible to exclude this possibility. Thus, GRB 180418A is simply the first \emph{candidate} SGRB showing reverse-shock emission in the optical. 

We have not been able to identify the host galaxy of SGRB 180418A (see Appendix~\ref{app:host}). If deep imaging were to conclusively identify the host and if a spectroscopic redshift were obtained, this might resolve the ambiguity about the nature of the burst. This absence emphasises the continuing need to obtain spectroscopy of bright GRBs.



%
%

\section*{acknowledgments}

We thank the referee for useful comments that helped to improve this article. We thank the staff of the Observatorio Astron\'omico Nacional on Sierra San Pedro M\'artir.
RATIR is a collaboration between the University of California, the Universidad Nacional Auton\'oma de M\'exico, NASA Goddard Space Flight Center, and Arizona State University, benefiting from the loan of an H2RG detector and hardware and software support from Teledyne Scientific and Imaging. RATIR, the automation of the Harold L. Johnson Telescope of the Observatorio Astron\'omico Nacional on Sierra San Pedro M\'artir, and the operation of both are funded through NASA grants NNX09AH71G, NNX09AT02G, NNX10AI27G, and NNX12AE66G, CONACyT grants INFR-2009-01-122785 and CB-2008-101958, UNAM PAPIIT grants IG100414, IA102917 and IA102019, UC MEXUS-CONACyT grant CN 09-283, and the Instituto de Astronom{\'\i}a of the Universidad Nacional Auton\'oma de M\'exico. We acknowledge the vital contributions of Neil Gehrels and Leonid Georgiev to the early development of RATIR.
TAROT has been built with the support of the Institut National des Sciences de l'Univers, CNRS, France. TAROT is funded by the CNES and thanks the help of the technical staff of the Observatoire de Haute Provence, OSU-Pytheas.

\clearpage
\renewcommand{\thesection}{\Alph{section}}
\setcounter{section}{0}

\section{Search for the GRB host galaxy}
\label{app:host}

\begin{figure}
\centering
 \includegraphics[width=0.5\textwidth]{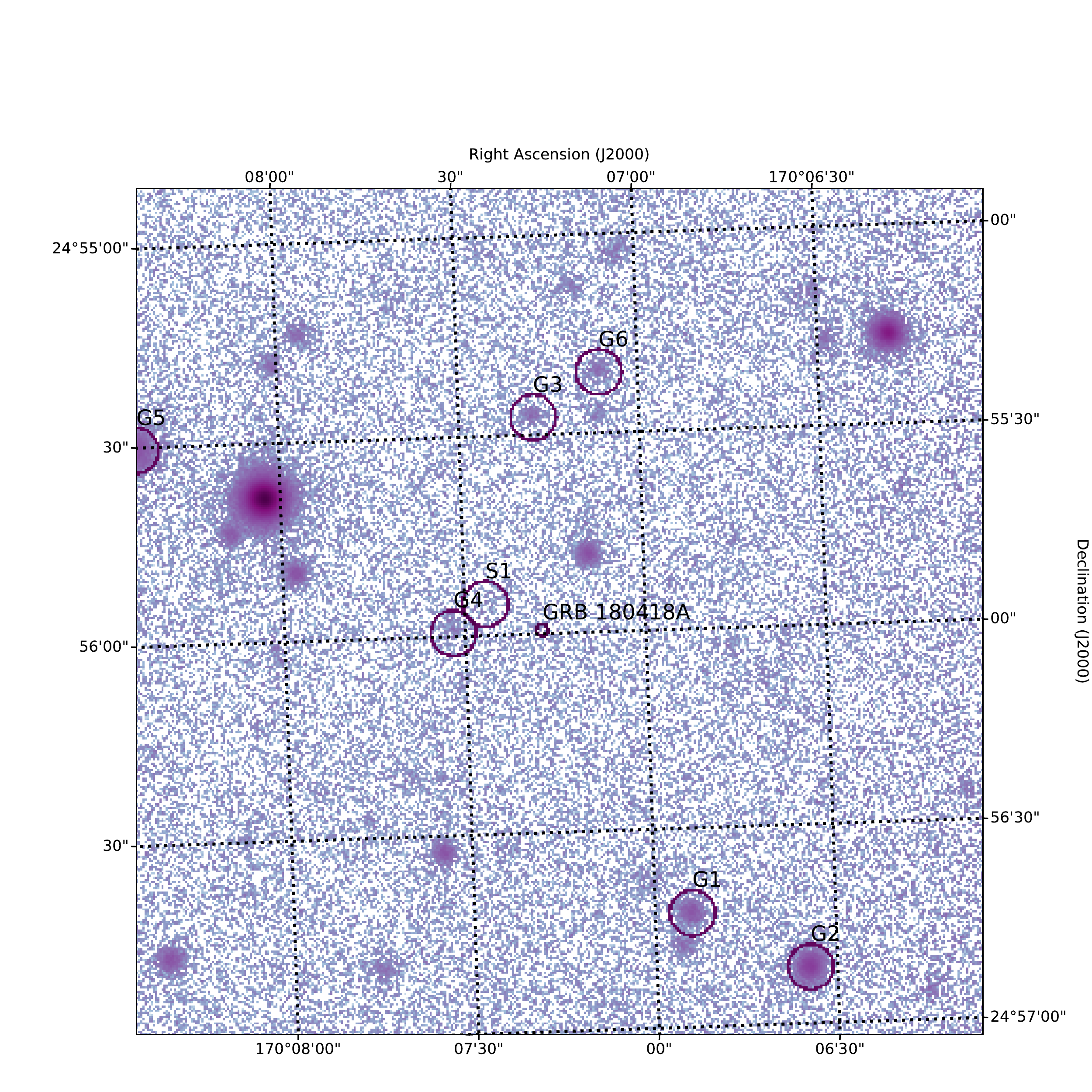}
 \caption{A deep RATIR $r$ image of the neighborhood of GRB 180418A. Information on the detected galaxies is given in Table~\ref{tab:host}.}
 \label{fig:host}
\end{figure}


We obtained deep photometry of the field with RATIR on the fourth night, at about $T+72$ hours. Our $r$-band image is shown in Figure~\ref{tab:host} and photometry of sources is given in Table~\ref{tab:host}. The first column is the object label as shown in Figure~\ref{tab:host}, the second is the SDSS DR12 ID, the third and fourth columns are the J2000 coordinates, the fifth and sixth columns are $r$ and $i$ AB magnitude, respectively, the seventh column is the angular separation $\theta$ in arcseconds, and the last column is the probability of a chance alignment $P(<\!\delta R)$ \citep{2002AJ....123.1111B}.


We detect no source at the position of the GRB afterglow, thus any coincident host galaxy must be fainter than our $3\sigma$ limits of $r> 24.0$ and $i>23.9$. Such a  faint galaxy would be consistent with our limit $z<1.3$ on the GRB redshift and,
if compared to field galaxies \citep{2007ApJ...664.1000B}, might suggest $z>0.3$.


\cite{22659} noted that there were no extended sources within 30 arcsec of the afterglow position down to a limit of $g>23.4$ and $r>22.5$. 
Thanks to our deeper photometry, we detect fainter sources, such as one at $r = 24.0$ and $i=23.7$ at a separation of 9 arcsec from the afterglow. Our images have a FWHM of about 2 arcsec, so we can say little about whether these faint sources are extended or not, although at these magnitudes we expect the majority to be galaxies \citep{2001AJ....122.1104Y}.



The two galaxies with the smallest chance probabilities are G4 with $P(<\!\delta R)=0.67$ and S1 with $P(<\!\delta R)=0.69$. 
G4 appears in the SDSS DR12 catalog with a photometric redshift of $z=0.80\pm0.12$ and an absolute magnitude of $M_r=-21.23$. If it were the host galaxy, the projected distance to the GRB would be 325 kpc. S1 is not in the SDSS DR12 catalog, but if it were at a typical SGRB redshift of 0.5 \citep{2014ARA&A..52...43B}, the projected offset to the GRB would be 137 kpc. Both of these projected distances are large compared to the sample of \cite{2008MNRAS.385L..10T} and \cite{2014ARA&A..52...43B}, whose offsets range up to 75 kpc. Thus, based both on the large chance probabilities $P(<\!\delta R)$ and the large offset distances, neither of these galaxies are likely to be the host of the GRB.

\clearpage
\LongTables

\begin{deluxetable*}{crrcc}
\tablecaption{UVOT observations of GRB 180418A\label{tab:datosuvot}}
\tablehead{\colhead{Filter}& \colhead{$t_\mathrm{i}$ (s)}& \colhead{$t_\mathrm{f}$ (s)}&\colhead{AB} &\colhead{SNR}}
\startdata
$v$ & 3242.6 & 4834.19 & $> 20.25$ & --- \\
$b$ & 4063.89 & 4263.66 & $20.43 \pm 0.30$ & 3.687 \\
$u$ & 3858.83 & 4058.61 & $20.52 \pm 0.20$ & 5.675 \\
$uvw1$ & 3654.22 & 3854.01 & $20.77 \pm0.25$ & 4.438 \\
$uvm2$ & 3449.18 & 3648.95 & $20.85 \pm0.30$ & 3.734 \\
$uvw2$ & 4475.76 & 4675.53 & $> 20.90$ & --- \\
White & 3086.17 & 3235.93 & $19.99 \pm 0.13$ & 9.123 \\
White & 4270.14 & 4469.9 & $20.40 \pm 0.15$ & 8.060
\enddata
\end{deluxetable*}

\begin{deluxetable*}{rrrcc}
\tablecaption{TAROT observations of GRB 180418A\label{tab:datostarot}}
\tablehead{\colhead{$t_\mathrm{i}$ (s)}& \colhead{$t_\mathrm{f}$ (s)}&\colhead{$r$}}
\startdata
28.0 & 36.0 & 14.30 $\pm$ 0.04 \\
36.0 & 43.0 & 14.24 $\pm$ 0.03 \\
43.0 & 51.0 & 14.40 $\pm$ 0.03 \\
51.0 & 58.0 & 14.61 $\pm$ 0.04 \\
58.0 & 66.0 & 14.98 $\pm$ 0.05 \\
66.0 & 73.0 & 15.42 $\pm$ 0.06 \\
73.0 & 81.0 & 15.52 $\pm$ 0.09 \\
81.0 & 88.0 & 15.83 $\pm$ 0.15 \\
100.0 & 130.0 & 15.96 $\pm$ 0.06 \\
141.0 & 171.0 & 16.50 $\pm$ 0.11 \\
181.0 & 211.0 & 16.58 $\pm$ 0.13 \\
221.0 & 251.0 & 17.10 $\pm$ 0.31 \\
262.0 & 292.0 & 16.90 $\pm$ 0.25 \\
302.0 & 392.0 & 17.00 $\pm$ 0.29
\enddata
\end{deluxetable*}

\clearpage

\begin{deluxetable*}{rrrcc}
\tablecaption{RATIR observations of GRB 180418A\label{tab:datosratir}}
\tablehead{\colhead{$t_\mathrm{i}$ (s)}& \colhead{$t_\mathrm{f}$ (s)}& \colhead{$t_\mathrm{exp}$ (s)}&\colhead{$r$}&\colhead{$i$}}
\startdata
120.6 & 200.6 & 80 & 16.39 $\pm$ 0.01 & 16.29 $\pm$ 0.01 \\
230.6 & 310.6 & 80 & 17.21 $\pm$ 0.01 & 16.96 $\pm$ 0.02 \\
322.4 & 402.4 & 80 & 17.56 $\pm$ 0.01 & 17.40 $\pm$ 0.01 \\
414.3 & 494.3 & 80 & 17.85 $\pm$ 0.02 & 17.93 $\pm$ 0.03 \\
521.6 & 601.6 & 80 & 18.13 $\pm$ 0.04 & 18.53 $\pm$ 0.06 \\
616.2 & 696.2 & 80 & 18.17 $\pm$ 0.02 & \nodata \\
617.1 & 697.1 & 80 & \nodata & 18.04 $\pm$ 0.03 \\
714.1 & 794.1 & 80 & 18.25 $\pm$ 0.04 & 18.27 $\pm$ 0.04 \\
812.1 & 892.1 & 80 & 18.43 $\pm$ 0.02 & 19.33 $\pm$ 0.05 \\
900.7 & 980.7 & 80 & 18.71 $\pm$ 0.05 & 18.89 $\pm$ 0.07 \\
999.6 & 1079.6 & 80 & 18.65 $\pm$ 0.06 & 18.50 $\pm$ 0.07 \\
1087.7 & 1167.7 & 80 & 18.66 $\pm$ 0.03 & 18.51 $\pm$ 0.04 \\
1199.0 & 1279.0 & 80 & 18.86 $\pm$ 0.03 & \nodata \\
1199.1 & 1279.0 & 80 & \nodata & 18.79 $\pm$ 0.04 \\
1287.2 & 1367.2 & 80 & \nodata & 18.75 $\pm$ 0.04 \\
1287.6 & 1367.6 & 80 & 18.84 $\pm$ 0.03 & \nodata \\
1380.3 & 1460.3 & 80 & 18.81 $\pm$ 0.03 & 18.83 $\pm$ 0.05 \\
1472.1 & 1552.1 & 80 & \nodata & 18.77 $\pm$ 0.05 \\
1570.0 & 1650.0 & 80 & 19.16 $\pm$ 0.06 & 18.94 $\pm$ 0.06 \\
1683.4 & 1763.3 & 80 & 19.17 $\pm$ 0.04 & \nodata \\
1684.3 & 1764.3 & 80 & \nodata & 19.17 $\pm$ 0.05 \\
1777.3 & 1857.3 & 80 & 19.22 $\pm$ 0.06 & 19.01 $\pm$ 0.06 \\
1883.6 & 1963.6 & 80 & 19.08 $\pm$ 0.04 & 18.93 $\pm$ 0.04 \\
1975.8 & 2055.8 & 80 & 19.36 $\pm$ 0.05 & 19.26 $\pm$ 0.06 \\
2064.0 & 2144.0 & 80 & 19.11 $\pm$ 0.08 & 19.06 $\pm$ 0.08 \\
2164.6 & 2244.6 & 80 & \nodata & 19.21 $\pm$ 0.07 \\
2164.9 & 2244.9 & 80 & 19.18 $\pm$ 0.05 & \nodata \\
2284.4 & 2364.4 & 80 & 19.46 $\pm$ 0.07 & 19.16 $\pm$ 0.07 \\
2377.5 & 2457.5 & 80 & 19.34 $\pm$ 0.05 & 19.53 $\pm$ 0.07 \\
2466.0 & 2546.0 & 80 & 19.74 $\pm$ 0.14 & 19.51 $\pm$ 0.15 \\
2558.5 & 2638.5 & 80 & 19.53 $\pm$ 0.06 & \nodata \\
2559.5 & 2639.5 & 80 & \nodata & 19.35 $\pm$ 0.06 \\
2655.9 & 2735.9 & 80 & 19.53 $\pm$ 0.08 & 19.20 $\pm$ 0.07 \\
2755.1 & 2835.1 & 80 & 19.97 $\pm$ 0.07 & 19.45 $\pm$ 0.06 \\
2842.3 & 2922.3 & 80 & 19.73 $\pm$ 0.08 & 19.49 $\pm$ 0.08 \\
2934.2 & 3014.2 & 80 & 19.63 $\pm$ 0.06 & 19.50 $\pm$ 0.08 \\
3023.6 & 3103.6 & 80 & 19.84 $\pm$ 0.10 & 19.64 $\pm$ 0.10 \\
3134.3 & 3214.3 & 80 & 19.92 $\pm$ 0.09 & \nodata \\
3135.3 & 3215.3 & 80 & \nodata & 19.64 $\pm$ 0.08 \\
3247.5 & 3327.5 & 80 & 19.90 $\pm$ 0.08 & 19.88 $\pm$ 0.09 \\
3341.3 & 3421.3 & 80 & 19.91 $\pm$ 0.11 & 19.73 $\pm$ 0.13 \\
3428.6 & 3508.6 & 80 & 19.67 $\pm$ 0.07 & 19.67 $\pm$ 0.08 \\
3521.2 & 3601.2 & 80 & 19.96 $\pm$ 0.12 & 19.77 $\pm$ 0.12 \\
3608.5 & 3688.5 & 80 & 19.80 $\pm$ 0.10 & 20.13 $\pm$ 0.14 \\
3714.2 & 3794.2 & 80 & 19.84 $\pm$ 0.09 & 19.80 $\pm$ 0.10 \\
3812.2 & 3892.2 & 80 & 19.87 $\pm$ 0.12 & 19.80 $\pm$ 0.13 \\
3910.9 & 3990.9 & 80 & 20.01 $\pm$ 0.12 & 19.81 $\pm$ 0.10 \\
4090.6 & 4170.6 & 80 & 20.05 $\pm$ 0.13 & 19.69 $\pm$ 0.10 \\
4207.0 & 4287.0 & 80 & 20.34 $\pm$ 0.15 & 19.70 $\pm$ 0.09 \\
4387.7 & 4467.7 & 80 & 20.30 $\pm$ 0.13 & 19.99 $\pm$ 0.10 \\
4479.2 & 4559.2 & 80 & 20.25 $\pm$ 0.12 & 20.06 $\pm$ 0.12 \\
4567.9 & 4647.9 & 80 & 20.34 $\pm$ 0.13 & 19.97 $\pm$ 0.11 \\
4666.5 & 4746.5 & 80 & 20.12 $\pm$ 0.11 & 20.08 $\pm$ 0.11 \\
4778.2 & 6292.3 & 1280 & 20.33 $\pm$ 0.04 & 20.20 $\pm$ 0.04 \\
4888.0 & 4968.1 & 80 & 19.96 $\pm$ 0.10 & \nodata \\
4888.1 & 4968.1 & 80 & \nodata & 19.89 $\pm$ 0.11 \\
4976.1 & 5056.1 & 80 & 20.28 $\pm$ 0.15 & \nodata \\
5068.0 & 5148.0 & 80 & 20.30 $\pm$ 0.13 & 20.29 $\pm$ 0.13 \\
5156.0 & 5236.0 & 80 & 20.36 $\pm$ 0.13 & 20.34 $\pm$ 0.14 \\
5247.9 & 5327.9 & 80 & 20.32 $\pm$ 0.13 & 20.04 $\pm$ 0.12 \\
5348.4 & 5428.4 & 80 & 19.92 $\pm$ 0.09 & 20.16 $\pm$ 0.13 \\
5447.4 & 5527.4 & 80 & 20.48 $\pm$ 0.16 & 20.35 $\pm$ 0.14 \\
5630.6 & 5710.6 & 80 & 20.47 $\pm$ 0.15 & 20.21 $\pm$ 0.14 \\
5728.5 & 5808.5 & 80 & 20.23 $\pm$ 0.13 & 20.29 $\pm$ 0.15 \\
5826.9 & 5906.9 & 80 & 20.34 $\pm$ 0.14 & 20.24 $\pm$ 0.13 \\
6113.6 & 6193.6 & 80 & 20.40 $\pm$ 0.15 & 20.34 $\pm$ 0.17 \\
6212.3 & 6292.3 & 80 & 20.43 $\pm$ 0.16 & \nodata \\
6324.4 & 7883.8 & 1280 & 20.72 $\pm$ 0.06 & 20.30 $\pm$ 0.12 \\
6324.4 & 7883.8 & 1280 & 20.72 $\pm$ 0.06 & 20.51 $\pm$ 0.05 \\
6435.7 & 6515.7 & 80 & \nodata & 20.24 $\pm$ 0.13 \\
6523.2 & 6603.2 & 80 & 20.44 $\pm$ 0.14 & 20.20 $\pm$ 0.12 \\
6623.3 & 6703.4 & 80 & 20.42 $\pm$ 0.14 & 20.67 $\pm$ 0.17 \\
6732.1 & 6812.1 & 80 & 20.61 $\pm$ 0.16 & 20.99 $\pm$ 0.24 \\
6825.8 & 6905.8 & 80 & 20.43 $\pm$ 0.14 & 20.25 $\pm$ 0.14 \\
6923.6 & 7003.6 & 80 & \nodata & 20.46 $\pm$ 0.17 \\
7111.3 & 7191.3 & 80 & \nodata & 20.48 $\pm$ 0.17 \\
7318.6 & 7398.6 & 80 & \nodata & 20.53 $\pm$ 0.18 \\
7418.1 & 7498.1 & 80 & \nodata & 20.52 $\pm$ 0.15 \\
8362.3 & 9887.7 & 1280 & 21.56 $\pm$ 0.15 & \nodata \\
8362.4 & 9887.7 & 1280 & \nodata & 20.85 $\pm$ 0.08 \\
8542.3 & 8622.3 & 80 & \nodata & 20.64 $\pm$ 0.25 \\
9940.3 & 11481.0 & 1280 & \nodata & 20.96 $\pm$ 0.10 \\
11513.1 & 13094.9 & 1280 & \nodata & 21.22 $\pm$ 0.13

\enddata
\end{deluxetable*}

\clearpage

\begin{deluxetable}{lcr}
\tabletypesize{\scriptsize}
\tablecaption{Fitting parameters of GRB 180418A \label{tab:fit}}
\tablehead{\colhead{}&\colhead{Parameter}&{Value}}
\startdata
Reverse shock&&\\
\hline
Electron index&$p$&2.35$\pm$0.03\\
&$t_\mathrm{\gamma}$&$35\pm 1$ s\\
$t<t_\mathrm{\gamma}$&$\alpha_\mathrm{reverse,R}$&$-5.56\pm 0.01$\\
$t> t_\mathrm{\gamma}$&$\alpha_\mathrm{reverse,D}$&$2.02\pm 0.01$\\[2ex]
\hline
Forward shock&&\\
\hline
Electron index&$p$&2.35$\pm$0.03\\
Optical&$\alpha_\mathrm{forward,optical}$&$1.01\pm0.01$\\
$X$-rays&$\alpha_\mathrm{forward,X}$&$1.04\pm0.05$\\
&$\beta$&$0.80\pm0.04$
\tablecomments{The variable $t_\mathrm{\gamma}$ is the time when is reached a maximum for reverse shock component. The temporal indices $\alpha$ are labeled with suffixes for reverse and forward and with $R$ or $D$ to refer to the rise and decay phase of the reverse component.}
\end{deluxetable}

\begin{deluxetable}{llr}
\tabletypesize{\scriptsize}
\tablecaption{Parameters found of the early and late afterglow of GRB 180418A \label{tab:fit2}}
\tablehead{\colhead{Parameter}&\colhead{Symbol}&{Value}}\\
\startdata
ISM density& $n$&0.15cm$^{-3}$\\
Lorentz Factor& $\Gamma$&160\\
Magnetization rate&$R_{B}=\left (B_{r}/B_{f} \right )$&14\\

&$\epsilon_{B,f}$&$10^{-3}$\\
&$\epsilon_{B,r}$&0.2\\
&$\epsilon_e$&0.1\\
Isotropic energy&$E_{iso}$&$0.77\times 10^{51}$ erg\\
Redshift &z&$0.5$\\
Optical luminosity&$L_{opt,7}$& $1.51\times 10^{42}$erg s$^{-1}$\\
Opening angle&$\theta_j$&\textgreater$ 7^\circ$\\
\enddata
\tablecomments{Most of these parameters are calculated using the work of \cite{2017ApJ...848...15F} and using the model of \cite{2000ApJ...545..807K}. The value of the redshift z=0.5 is a estimation according with the average value in SGRBs and, from it, we calculate the isotropic kinetic energy $E_{iso}$, $L_{opt,7}$ and 
the limit value of $\theta_j$}
\end{deluxetable}

\begin{deluxetable}{lrrcr}
\tabletypesize{\scriptsize}
\tablewidth{0pt}
\tablecaption{GRBs with Signature of a Reverse Shock.\label{tab:list}}
\tablehead{\colhead{GRB}&\colhead{z}&\colhead{T$_{90}$}&\colhead{Evidence}&\colhead{Reference}}\\
\startdata
GRB 990123&1.60&63.3&B&1, 2, 3\\
GRB 021004 &2.33&100.0&A&4, 5\\ 
GRB 021211 &0.80&2.67&B&6\\ 
GRB 050525A &0.61&8.8&C&7, 8, 9, 10\\ 
GRB 050904 &6.9&225.0&C&11,12\\ 
GRB 060111B&1-2&59&C&13, 14\\ 
GRB 060117&0.45&16.0&C&15, 16, 17\\ 
GRB 060908 &1.88&19.30&A&18\\ 
GRB 061126 &1.56&191&B&19, 20\\ 
GRB 080319B&0.94&$>$50&A&21, 22\\ 
GRB 081007 &0.53&8.0&B&23, 24\\ 
GRB 090102 &1.55&27.0&A&18, 25, 26\\ 
GRB 090424 &0.54&48.0&A&23,27\\ 
GRB 090902B &1.82&21.0&C&28, 29, 30\\ 
GRB 091024 &1.1&1020&C&31, 32\\ 
GRB 110205A &2.11&257&C&33, 34, 35\\ 
GRB 130427A &0.34&163&B&36, 37, 38\\ 
GRB 160625B &1.41&35&B&39, 40 \\ 
GRB 180418A & $<1.31$ &1.5&B& This work
\enddata
\tablenotetext{A}{Confirmed.}
\tablenotetext{B}{Strongly confirmed.}
\tablenotetext{C}{There are another possibilities but due to the
lack of good early-time photometric coverage.}
\tablerefs{
(1) \cite{1999MNRAS.306L..39M}; (2) \cite{1999ApJ...518L...1B}; (3)
\cite{224}; (4) 
\cite{2010A&A...517A..61C}, (5) \cite{2003ApJ...582L..75K}, (6) \cite{2003A&A...402L...9W}, (7) \cite{2005ApJ...633.1027S}, (8) \cite{2005GCN..3479....1C}, (9) \cite{2005GCN..3483....1F}, (10) \cite{2005GCN..3467....1M}, (11)
\cite{3938}, (12)
\cite{2006ApJ...636L..69W}, (13)
\cite{2006A&A...451L..39K}, (14) \cite{2009A&A...503..783S}, (15)
\cite{2006NCimB.121.1495J}, (16) 
\cite{4544}, (17) 
\cite{4538}, (18) 
\cite{2014ApJ...785...84J}, (19)
\cite{2008ApJ...687..443G}, (20) \cite{2008ApJ...672..449P}, (21)
\cite{2009ApJ...692.1662Y}, (22)
\cite{2008GCN..7444....1V}, (23)
\cite{2013ApJ...774..114J}, (24)
\cite{2011A&A...525A.109D}, (25) \cite{2009GCN..8766....1D}, (26),
\cite{2010MNRAS.405.2372G}, (27)
\cite{2009GCN..9231....1S}, (28) \cite{2010ApJ...714..799P}, (29)
\cite{2009GCN..9873....1C}, (30)
\cite{2009GCN..9872....1D}, (31)
\cite{2011A&A...528A..15G}, (32) \cite{2011ApJ...743..154C}, (33) \cite{2011GCN.11635....1D}, (34) \cite{2011GCN.11646....1M}, (35) \cite{2012ApJ...748...59G}, (36) \cite{2014ApJ...781...37P}, (37)
\cite{2013GCN.14470....1B}, (38) \cite{2013GCN.14455....1L}, (39) \cite{2017Natur.547..425T}, (40) \cite{Zhang18}}
\end{deluxetable}

\begin{deluxetable*}{rcccrrrrrr}
\tabletypesize{\scriptsize}
\tablewidth{0pt}
\tablecaption{Candidate galaxies to be the host galaxy of GRB 180418A\label{tab:host}}
\tablehead{\colhead{Label}&\colhead{SDSS ID}&\colhead{RA}&\colhead{DEC}&\colhead{$r$}&\colhead{$i$}&\colhead{$\delta R$ ['']}&\colhead{$P(<\delta R)$}}\\
\startdata
G1 & 1237667551421006080 & 170.1150 & 24.9451 & 21.1390 $\pm$ 0.0274 & 20.8640 $\pm$ 0.0240 & 48.18 & 0.997\\
G2 & 1237667551421006592 & 170.1097 & 24.9475 & 20.0264 $\pm$ 0.0115 &19.5144 $\pm$ 0.0084 & 64.84 & 0.986\\
G3 & 1237667551421006592 & 170.1216 & 24.9242 & 22.2155 $\pm$ 0.0714 &21.9499 $\pm$ 0.0613 & 32.11 & 0.996\\
G4 & 1237667551421006848 & 170.1256 & 24.9331 & 22.7085 $\pm$ 0.1149 &22.5791$\pm$ 0.1089 & 13.39 & 0.674\\
G5 & 1237667551421006592 & 170.1400 & 24.9250 & 20.2362 $\pm$ 0.0141 &19.7060 $\pm$ 0.0099 & 66.93 & 1.000\\
G6 & 1237667551421006592 & 170.1185 & 24.9224 & 22.1486 $\pm$ 0.0662 &21.7405 $\pm$ 0.0509 & 39.83 & 0.999\\
S1&\nodata 			       & 170.1241 & 24.9319 &23.9651 $\pm$ 0.0000 &23.7360 $\pm$ 0.3158 & 9.45 & 0.690

\enddata
\end{deluxetable*}


\begin{thebibliography}{200}
\bibitem[Abbott et al.(2017a)]{2017PhRvL.119p1101A} Abbott, B.~P., Abbott, R., Abbott, T.~D., et al.\ 2017, Physical Review Letters, 119, 161101
\bibitem[Abbott et al.(2017b)]{2017ApJ...848L..12A} Abbott, B.~P., Abbott, R., Abbott, T.~D., et al.\ 2017, \apjl, 848, L12
\bibitem[Akerlof et al.(1999)]{1999Natur.398..400A} Akerlof, C., Balsano, R., Barthelmy, S., et al.\ 1999, \nat, 398, 400
\bibitem[Amati et al.(2008)]{2008MNRAS.391..577A} Amati, L., Guidorzi, C., Frontera, F., et al.\ 2008, \mnras, 391, 577 
\bibitem[Arnaud(1996)]{1996ASPC..101...17A} Arnaud, K.~A.\ 1996, Astronomical Data Analysis Software and Systems V, 101, 17
\bibitem[Barthelmy et al.(2013)]{2013GCN.14470....1B} Barthelmy, S.~D., Baumgartner, W.~H., Cummings, J.~R., et al.\ 2013, GRB Coordinates Network, Circular Service, No.~14470, \#1 (2013), 14470, 1 
\bibitem[Berger et al.(2007)]{2007ApJ...664.1000B} Berger, E., Fox, D.~B., Price, P.~A., et al.\ 2007, \apj, 664, 1000 
\bibitem[Berger(2010)]{2010ApJ...722.1946B} Berger, E.\ 2010, \apj, 722, 1946 \bibitem[Berger(2014)]{2014ARA&A..52...43B} Berger, E.\ 2014, \araa, 52, 43 
\bibitem[Bissaldi \& Veres(2018)]{22656} Bissaldi, E., \& Veres, P.\ 2018, GRB Coordinates Network, Circular Service, No.~22656, \#1 (2018/April-0), 22656, 1 
\bibitem[Blandford \& McKee(1976)]{1976PhFl...19.1130B} Blandford, R.~D., \& McKee, C.~F.\ 1976, Physics of Fluids, 19, 1130 
\bibitem[Bloom et al.(1999)]{1999ApJ...518L...1B} Bloom, J.~S., Odewahn, S.~C., Djorgovski, S.~G., et al.\ 1999, \apjl, 518, L1 
\bibitem[Bloom et al.(2002)]{2002AJ....123.1111B} Bloom, J.~S., Kulkarni, S.~R., \& Djorgovski, S.~G.\ 2002, \aj, 123, 1111
\bibitem[Bloom et al.(2009)]{2009ApJ...691..723B} Bloom, J.~S., Perley, D.~A., Li, W., et al.\ 2009, \apj, 691, 723
\bibitem[Breeveld et al.(2011)]{2011AIPC.1358..373B} Breeveld, A.~A., Landsman, W., Holland, S.~T., et al.\ 2011, American Institute of Physics Conference Series, 1358, 373 
\bibitem[Bromberg et al.(2013)]{2013ApJ...764..179B} Bromberg, O., Nakar, E., Piran, T., et al.\ 2013, \apj, 764, 179.
\bibitem[Brun \& Rademakers(1997)]{1997NIMPA.389...81B} Brun, R., \& Rademakers, F.\ 1997, Nuclear Instruments and Methods in Physics Research A, 389, 81 
\bibitem[Burrows et al.(2006)]{2006ApJ...653..468B} Burrows, D.~N., Grupe, D., Capalbi, M., et al.\ 2006, \apj, 653, 468
 \bibitem[Butler(2007)]{2007AJ....133.1027B} Butler, N.~R.\ 2007, \aj, 133, 1027 
 \bibitem[Butler et al.(2007)]{2007ApJ...671..656B} Butler, N.~R., Kocevski, D., Bloom, J.~S., \& Curtis, J.~L.\ 2007, \apj, 671, 656 
\bibitem[Butler et al.(2012)]{2012SPIE.8446E..10B} Butler, N., Klein, C., Fox, O., et al.\ 2012, \procspie, 8446, 844610
\bibitem[Cano(2013)]{2013MNRAS.434.1098C} Cano, Z.\ 2013, \mnras, 434, 1098.
\bibitem[Cardelli et al.(1989)]{1989ApJ...345..245C} Cardelli, J.~A., Clayton, G.~C., \& Mathis, J.~S.\ 1989, \apj, 345, 245
\bibitem[Castro-Tirado et al.(2010)]{2010A&A...517A..61C} Castro-Tirado, A.~J., M{\o}ller, P., Garc{\'{\i}}a-Segura, G., et al.\ 2010, \aap, 517, A61 
\bibitem[Choi et al.(2018)]{22668} Choi, C., Kim, Y., Park, W., Shin, S., \& Im, M.\ 2018, GRB Coordinates Network, Circular Service, No.~22668, \#1 (2018/April-0), 22668, 1 
\bibitem[Cucchiara et al.(2009)]{2009GCN..9873....1C} Cucchiara, A., Fox, D.~B., Tanvir, N., \& Berger, E.\ 2009, GRB Coordinates Network, 9873, 1 
\bibitem[Cucchiara et al.(2011)]{2011ApJ...743..154C} Cucchiara, A., Cenko, S.~B., Bloom, J.~S., et al.\ 2011, \apj, 743, 154 
\bibitem[Cummings et al.(2005)]{2005GCN..3479....1C} Cummings, J., Barbier, L., Barthelmy, S., et al.\ 2005, GRB Coordinates Network, 3479, 1 
\bibitem[Cummings et al.(2006)]{4538}Cummings, J., Barbier, L., Barthelmy, S., et al.\ 2006, GRB Coordinates Network, 4538, 1 
\bibitem[da Silva et al.(2011)]{2011GCN.11635....1D} da Silva, R., Fumagalli, M., Worseck, G., \& Prochaska, X.\ 2011, GRB Coordinates Network, Circular Service, No.~11635, \#1 (2011), 11635, 1 
\bibitem[D'Elia et al.(2018)]{22646} D'Elia, V., D'Ai, A., Evans, P.~A., et al.\ 2018, GRB Coordinates Network, Circular Service, No.~22646, \#1 (2018/April-0), 22646, 1 
\bibitem[De Pasquale et al.(2010)]{2010ApJ...709L.146D} De Pasquale, M., Schady, P., Kuin, N.~P.~M., et al.\ 2010, \apjl, 709, L146
\bibitem[de Ugarte Postigo et al.(2009)]{2009GCN..8766....1D} de Ugarte Postigo, A., Jakobsson, P., Malesani, D., et al.\ 2009, GRB Coordinates Network, 8766, 1 
\bibitem[de Ugarte Postigo et al.(2011)]{2011A&A...525A.109D} de Ugarte Postigo, A., Horv{\'a}th, I., Veres, P., et al.\ 2011, \aap, 525, A109 
\bibitem[Eichler et al.(1989)]{1989Natur.340..126E} Eichler, D., Livio, M., Piran, T., et al.\ 1989, \nat, 340, 126
 \bibitem[Evans et al.(2009)]{2009MNRAS.397.1177E} Evans, P.~A., Beardmore, A.~P., Page, K.~L., et al.\ 2009, \mnras, 397, 1177 
 \bibitem[Fan \& Wei(2005)]{2005MNRAS.364L..42F} Fan, Y.~Z., \& Wei, D.~M.\ 2005, \mnras, 364, L42 
\bibitem[Fan et al.(2005)]{2005ApJ...628L..25F} Fan, Y.~Z., Zhang, B., \& Wei, D.~M.\ 2005, \apjl, 628, L25 
\bibitem[Fenimore et al.(1999)]{1999ApJ...518L..73F} Fenimore, E.~E., Ramirez-Ruiz, E., \& Wu, B.\ 1999, \apjl, 518, L73 
\bibitem[Foley et al.(2005)]{2005GCN..3483....1F} Foley, R.~J., Chen, H.-W., Bloom, J., \& Prochaska, J.~X.\ 2005, GRB Coordinates Network, 3483, 1
\bibitem[Fong et al.(2012)]{2012ApJ...756..189F} Fong, W., Berger, E., Margutti, R., et al.\ 2012, \apj, 756, 189 
\bibitem[Fong et al.(2014)]{2014ApJ...780..118F} Fong, W., Berger, E., Metzger, B.~D., et al.\ 2014, \apj, 780, 118 
\bibitem[Fong et al.(2018)]{22659} Fong, W., Tanvir, N.~R., Levan, A.~J., \& Chornock, R.\ 2018, GRB Coordinates Network, Circular Service, No.~22659, \#1 (2018/April-0), 22659, 1  
\bibitem[Fraija et al.(2016)]{2016ApJ...818..190F} Fraija, N., Lee, W.~H. \& Veres, P., \ 2016, \apj, 818, 190
 \bibitem[Fraija et al.(2016)]{2016ApJ...831...22F} Fraija, N., Lee, W.~H., Veres, P., \& Barniol Duran, R.\ 2016, \apj, 831, 22
 \bibitem[Fraija et al.(2017)]{2017ApJ...848...94F} Fraija, N., Lee, W.~H., Araya, M., et al. \ 2017, \apj, 848, 94
 \bibitem[Fraija et al.(2017b)]{2017ApJ...848...15F} Fraija, N., Veres, P., Zhang, B.~B., et al.\ 2017, \apj, 848, 15
 \bibitem[Fraija et al.(2017c)]{2017arXiv170101184F} Fraija, N., Lee, W.~H., Veres, P., et al.\ 2017, arXiv e-prints , arXiv:1701.01184.

\bibitem[Fraija \& Veres(2018)]{2018ApJ...859...70F} Fraija, N. \& Veres, P. \ 2018, \apj, 859, 70
\bibitem[Fraija et al.(2019)]{2019ApJ...871..123F} Fraija, N., De Colle, F, Veres, P. et al. \ 2019, \apj, 871, 123
\bibitem[Fraija et al.(2019)]{2019ApJ...871..200F} Fraija, N., Pedreira, A.~C.~C.~d.~E.~S. and Veres, P. \ 2019, \apj, 871, 200
\bibitem[Gao et al.(2013)]{2013NewAR..57..141G} Gao, H., Lei, W.-H., Zou, Y.-C., Wu, X.-F., \& Zhang, B.\ 2013, New Astronomy Review, 57, 141
\bibitem[Gao \& M{\'e}sz{\'a}ros(2015)]{2015AdAst2015E..13G} Gao, H., \& M{\'e}sz{\'a}ros, P.\ 2015, Advances in Astronomy, 2015, 192383 
\bibitem[Gavazzi et al.(2010)]{2010A&A...517A..73G} Gavazzi, G., Fumagalli, M., Cucciati, O., \& Boselli, A.\ 2010, \aap, 517, A73 
\bibitem[Gehrels et al.(2006)]{2006Natur.444.1044G} Gehrels, N., Norris, J.~P., Barthelmy, S.~D., et al.\ 2006, \nat, 444, 1044.
\bibitem[Gehrels, \& Razzaque(2013)]{2013FrPhy...8..661G} Gehrels, N., \& Razzaque, S.\ 2013, Frontiers of Physics, 8, 661.
\bibitem[Gendre et al.(2010)]{2010MNRAS.405.2372G} Gendre, B., Klotz, A., Palazzi, E., et al.\ 2010, \mnras, 405, 2372 
\bibitem[Gendre et al.(2012)]{2012ApJ...748...59G} Gendre, B., Atteia, J.~L., Bo{\"e}r, M., et al.\ 2012, \apj, 748, 59
\bibitem[Giacomazzo et al.(2011)]{2011PhRvD..83d4014G} Giacomazzo, B., Rezzolla, L. and Baiotti, L.\ 2011, \prd, 83, 044014
\bibitem[Goad et al.(2018)]{22650} Goad, M.~R., Osborne, J.~P., Beardmore, A.~P., \& Evans, P.~A.\ 2018, GRB Coordinates Network, Circular Service, No.~22650, \#1 (2018/April-0), 22650, 1 
\bibitem[Gomboc et al.(2008)]{2008ApJ...687..443G} Gomboc, A., Kobayashi, S., Guidorzi, C., et al.\ 2008, \apj, 687, 443
\bibitem[Gomboc et al.(2009)]{2009AIPC.1133..145G} Gomboc, A., Kobayashi, S., Mundell, C.~G., et al.\ 2009, American Institute of Physics Conference Series, 1133, 145 
\bibitem[Granot \& Sari(2002)]{2002ApJ...568..820G} Granot, J., \& Sari, R.\ 2002, \apj, 568, 820 
\bibitem[Granot(2005)]{2005ApJ...631.1022G} Granot, J. 2005, ApJ 631, 1022
\bibitem[Greiner et al.(2008)]{2008PASP..120..405G} Greiner, J., Bornemann, W., Clemens, C., et al.\ 2008, \pasp, 120, 405
\bibitem[Gruber et al.(2011)]{2011A&A...528A..15G} Gruber, D., Kr{\"u}hler, T., Foley, S., et al.\ 2011, \aap, 528, A15 
\bibitem[Guidorzi et al.(2018)]{22648} Guidorzi, C., Martone, R., Kobayashi, S., et al.\ 2018, GRB Coordinates Network, Circular Service, No.~22648, \#1 (2018/April-0), 22648, 1 
\bibitem[Hjorth et al.(2003)]{2003Natur.423..847H} Hjorth, J., Sollerman, J., M{\o}ller, P., et al.\ 2003, \nat, 423, 847 
\bibitem[Hogg et al.(1997)]{1997MNRAS.288..404H} Hogg, D.~W., Pahre, M.~A., McCarthy, J.~K., et al.\ 1997, \mnras, 288, 404.
\bibitem[Horiuchi et al.(2018)]{22670} Horiuchi, T., Hanayama, H., Honma, M., et al.\ 2018, GRB Coordinates Network, Circular Service, No.~22670, \#1 (2018/April-0), 22670, 1 
\bibitem[Japelj et al.(2014)]{2014ApJ...785...84J} Japelj, J., Kopa{\v c}, D., Kobayashi, S., et al.\ 2014, \apj, 785, 84 
\bibitem[Jel{\'{\i}}nek et al.(2006)]{2006NCimB.121.1495J} Jel{\'{\i}}nek, M., Prouza, M., Kub{\'a}nek, P., et al.\ 2006, Nuovo Cimento B Serie, 121, 1495 
\bibitem[Jin et al.(2013)]{2013ApJ...774..114J} Jin, Z.-P., Covino, S., Della Valle, M., et al.\ 2013, \apj, 774, 114 
\bibitem[Kann et al.(2010)]{2010ApJ...720.1513K} Kann, D.~A., Klose, S., Zhang, B., et al.\ 2010, \apj, 720, 1513 
\bibitem[Kippen et al.(1999)]{224} Kippen, R. et al. 1999, GRB Coordinates Network, 224, 1
\bibitem[Klotz et al.(2006)]{2006A&A...451L..39K} Klotz, A., Gendre, B., Stratta, G., et al.\ 2006, \aap, 451, L39 
\bibitem[Klotz et al.(2008)]{2008PASP..120.1298K} Klotz, A., Bo{\"e}2006ApJ...642..354Zr, M., Eysseric, J., et al.\ 2008, \pasp, 120, 1298
\bibitem[Klotz et al.(2018)]{22671} Klotz, A., Atteia, J.~L., Boer, M., Eymar, L., \& Gendre, B.\ 2018, GRB Coordinates Network, Circular Service, No.~22671, \#1 (2018/April-0), 22671, 1 
\bibitem[Kobayashi(2000)]{2000ApJ...545..807K} Kobayashi, S.\ 2000, \apj, 545, 807 
\bibitem[Kobayashi, \& Sari(2000)]{2000ApJ...542..819K} Kobayashi, S., \& Sari, R.\ 2000, \apj, 542, 819.\bibitem[Kobayashi \& Zhang(2003)]{2003ApJ...582L..75K} Kobayashi, S., \& Zhang, B.\ 2003, \apjl, 582, L75 
\bibitem[Kobayashi et al.(2007)]{2007ApJ...655..391K} Kobayashi, S., Zhang, B., M{\'e}sz{\'a}ros, P., \& Burrows, D.\ 2007, \apj, 655, 391 
\bibitem[Kouveliotou et al.(1993)]{Kouveliotou93} Kouveliotou, C., Meegan, C. A., Fishman, G. J., Bhat, N. P., Briggs, M. S., Koshut, T. M., Paciesas, W. S., \& Pendleton, G. N. 1993, ApJ, 413, L101
\bibitem[Kumar \& Zhang(2015)]{2015PhR...561....1K} Kumar, P., \& Zhang, B.\ 2015, \physrep, 561, 1 
\bibitem[Lang et al.(2010)]{2010AJ....139.1782L} Lang, D., Hogg, D.~W., Mierle, K., Blanton, M., \& Roweis, S.\ 2010, \aj, 139, 1782 
\bibitem[Perley et al.(2014)]{2014ApJ...781...37P} Perley, D.~A., Cenko, S.~B., Corsi, A., et al.\ 2014, \apj, 781, 37 
\bibitem[Lee \& Ramirez-Ruiz(2007)]{2007NJPh....9...17L} Lee, W.~H., \& Ramirez-Ruiz, E.\ 2007, New Journal of Physics, 9, 17 
\bibitem[Levan et al.(2013)]{2013GCN.14455....1L} Levan, A.~J., Cenko, S.~B., Perley, D.~A., \& Tanvir, N.~R.\ 2013, GRB Coordinates Network, Circular Service, No.~14455, \#1 (2013), 14455, 1 
\bibitem[Littlejohns et al.(2015)]{2015MNRAS.449.2919L} Littlejohns, O.~M., Butler, N.~R., Cucchiara, A., et al.\ 2015, \mnras, 449, 2919
\bibitem[Liu et al.(2018)]{22655} Liu, Z., Melandri, A., D'Avanzo, P., et al.\ 2018, GRB Coordinates Network, Circular Service, No.~22655, \#1 (2018/April-0), 22655, 1 
\bibitem[(Lloyd-Ronning 2018)]{2018Galax...6..103L} Lloyd-Ronning, N.\ 2018, Galaxies, 6, 468
\bibitem[Malesani et al.(2018)]{22660} Malesani, D., Heintz, K.~E., Stone, M., \& Stone, J.\ 2018, GRB Coordinates Network, Circular Service, No.~22660, \#1 (2018/April-0), 22660, 1 
\bibitem[Markwardt et al.(2005)]{2005GCN..3467....1M} Markwardt, C., Barthelmy, S., Barbier, L., et al.\ 2005, GRB Coordinates Network, 3467, 1 
\bibitem[Markwardt et al.(2011)]{2011GCN.11646....1M} Markwardt, C.~B., Barthelmy, S.~D., Baumgartner, W.~H., et al.\ 2011, GRB Coordinates Network, Circular Service, No.~11646, \#1 (2011), 11646, 1 
\bibitem[Meszaros \& Rees(1993)]{1993ApJ...405..278M} Meszaros, P., \& Rees, M.~J.\ 1993, \apj, 405, 278 
\bibitem[Meszaros and Rees(1997)]{1997ApJ...476..232M} Meszaros, P., Rees, M.~J.\ 1997.\ Optical and Long-Wavelength Afterglow from Gamma-Ray Bursts.\ The Astrophysical Journal 476, 232.
\bibitem[M{\'e}sz{\'a}ros \& Rees(1999)]{1999MNRAS.306L..39M} M{\'e}sz{\'a}ros, P., \& Rees, M.~J.\ 1999, \mnras, 306, L39
\bibitem[Misra et al.(2018)]{22663} Misra, K.  et al.\ 2018, GRB Coordinates Network, 22663, 1
\bibitem[Mundell et al.(2013)]{2013Natur.504..119M} Mundell, C.~G., Kopa{\v c}, D., Arnold, D.~M., et al.\ 2013, \nat, 504, 119
\bibitem[Narayan et al.(1992)]{1992ApJ...395L..83N} Narayan, R., Paczynski, B. and Piran, T.\ 1992, \apj, 395, L83
\bibitem[Narayana Bhat et al.(2016)]{2016ApJS..223...28N} Narayana Bhat, P., Meegan, C.~A., von Kienlin, A., et al.\ 2016, \apjs, 223, 28 
\bibitem[Nicuesa Guelbenzu et al.(2012)]{2012A&A...538L...7N} Nicuesa Guelbenzu, A., Klose, S., Kr{\"u}hler, T., et al.\ 2012, \aap, 538, L7 
\bibitem[Nicuesa Guelbenzu et al.(2011)]{2011A&A...531L...6N} Nicuesa Guelbenzu, A., Klose, S., Rossi, A., et al.\ 2011, \aap, 531, L6 
\bibitem[Norris et al.(2000)]{2000ApJ...534..248N} Norris, J.~P., Marani, G.~F., \& Bonnell, J.~T.\ 2000, \apj, 534, 248 
\bibitem[Norris(2002)]{2002ApJ...579..386N} Norris, J.~P.\ 2002, \apj, 579, 386
\bibitem[Osborne et al.(2018)]{22649} Osborne, J.~P., Burrows, D.~N., Kennea, J.~A., et al.\ 2018, GRB Coordinates Network, Circular Service, No.~22649, \#1 (2018/April-0), 22649, 1 
\bibitem[Paczynski \& Rhoads(1993)]{1993ApJ...418L...5P} Paczynski, B., \& Rhoads, J.~E.\ 1993, \apjl, 418, L5 
\bibitem[de Palma et al.(2009)]{2009GCN..9872....1D} de Palma, F., Bissaldi, E., Tajima, H., et al.\ 2009, GRB Coordinates Network, 9872, 1
\bibitem[Palmer et al.(2018)]{22658} Palmer, D.~M., Barthelmy, S.~D., Cummings, J.~R., et al.\ 2018, GRB Coordinates Network, Circular Service, No.~22658, \#1 (2018/April-0), 22658, 1 \bibitem[Pandey et al.(2010)]{2010ApJ...714..799P} Pandey, S.~B., Swenson, C.~A., Perley, D.~A., et al.\ 2010, \apj, 714, 799 
\bibitem[Pelangeon \& Atteia(2006)]{4544} Pelangeon, A. \& Atteia, J. L.\ 2006, GRB Coordinates Network, 4544,1
\bibitem[Peng et al.(2005)]{2005ApJ...626..966P} Peng, F., K{\"o}nigl, A., \& Granot, J.\ 2005, \apj, 626, 966 
\bibitem[Perley et al.(2008)]{2008ApJ...672..449P} Perley, D.~A., Bloom, J.~S., Butler, N.~R., et al.\ 2008, \apj, 672, 449 
\bibitem[Piro et al.(2017)]{2017ApJ...844L..19P} Piro, A.~L., Giacomazzo, B., \& Perna, R.\ 2017, \apj, 844, L19.
\bibitem[Planck Collaboration et 
al.(2014)]{2014A&A...571A...1P} Planck Collaboration, Ade, P.~A.~R., Aghanim, N., et al.\ 2014, \aap, 571, A1 
\bibitem[Poole et al.(2008)]{2008MNRAS.383..627P} Poole, T.~S., Breeveld, A.~A., Page, M.~J., et al.\ 2008, \mnras, 383, 627
\bibitem[Piran (1999)]{1999PhR...314..575P} Piran, T.\ 1999, \physrep, 314, 575
\bibitem[Racusin et al.(2008)]{2008Natur.455..183R} Racusin, J.~L. and Karpov, S.~V. and Sokolowski, M. and Granot, J. and Wu, X. ~F., et al.\ 2008, \nat, 455, 183 
\bibitem[Rees \& M{\'e}sz{\'a}ros(1998)]{1998ApJ...496L...1R} Rees, M.~J., \& M{\'e}sz{\'a}ros, P.\ 1998, \apjl, 496, L1
\bibitem[Rosswog \& Ramirez-Ruiz(2002)]{2002MNRAS.336L...7R} Rosswog, S. and Ramirez-Ruiz, E.\ 2002, \mnras, 336, L7
\bibitem[Ruffert \& Janka(1998)]{1998A&A...338..535R} Ruffert, M. and Janka, H.~T.\ 1998, \aap, 338, 535
\bibitem[Sakamoto et al.(20005)]{3938} Sakamoto, W., et al.\ 2005, GRB Coordinates Network, 3938, 1
\bibitem[Sakamoto et al.(2009)]{2009GCN..9231....1S} Sakamoto, T., Barthelmy, S.~D., Baumgartner, W.~H., et al.\ 2009, GRB Coordinates Network, 9231, 1 
\bibitem[Sari et al.(1998)]{1998ApJ...497L..17S} Sari, R., Piran, T., \& Narayan, R.\ 1998, \apjl, 497, L17 
\bibitem[Sari \& Piran(1999a)]{1999ApJ...517L.109S} Sari, R., \& Piran, T.\ 1999, \apjl, 517, L109 
\bibitem[Sari \& Piran(1999b)]{1999ApJ...520..641S} Sari, R., \& Piran, T.\ 1999, \apj, 520, 641 
\bibitem[Sari et al.(1999)]{1999ApJ...519L..17S} Sari, R., Piran, T., \& Halpern, J.~P.\ 1999, \apjl, 519, L17 
\bibitem[Schady(2018)]{22662} Schady, P.\ 2018, GRB Coordinates Network, Circular Service, No.~22662, \#1 (2018/April-0), 22662, 1 \bibitem[Schady \& Chen(2018)]{22666} Schady, P., \& Chen, T.-W.\ 2018, GRB Coordinates Network, Circular Service, No.~22666, \#1 (2018/April-0), 22666, 1 
\bibitem[Shao \& Dai(2005)]{2005ApJ...633.1027S} Shao, L., \& Dai, Z.~G.\ 2005, \apj, 633, 1027 
\bibitem[Siegel \& D'Elia(2018)]{22665} Siegel, M.~H., \& D'Elia, V.\ 2018, GRB Coordinates Network, Circular Service, No.~22665, \#1 (2018/April-0), 22665, 1 
\bibitem[Soderberg \& Ramirez-Ruiz(2002)]{2002MNRAS.330L..24S} Soderberg, A.~M., \& Ramirez-Ruiz, E.\ 2002, \mnras, 330, L24 
\bibitem[Sota et al.(2018)]{22657} Sota, A., Hu, Y., Tello, J.~C., Carrasco, I., \& Castro-Tirado, A.~J.\ 2018, GRB Coordinates Network, Circular Service, No.~22657, \#1 (2018/April-0), 22657, 1 
\bibitem[Steele et al.(2009)]{2009Natur.462..767S} Steele, I.~A., Mundell, C.~G., Smith, R.~J., et al.\ 2009, \nat, 462, 767
\bibitem[Steidel \& Hamilton(1993)]{1993AJ....105.2017S} Steidel, C.~C., \& Hamilton, D.\ 1993, \aj, 105, 2017
\bibitem[Stratta et al.(2007)]{2007A&A...474..827S} Stratta, G., D'Avanzo, P., Piranomonte, S., et al.\ 2007, \aap, 474, 827
\bibitem[Stratta et al.(2008)]{2008AIPC.1000..297S} Stratta, G., D'Avanzo, P., Piranomonte, S., et al.\ 2008, American Institute of Physics Conference Series, 1000, 297 
\bibitem[Stratta et al.(2009)]{2009A&A...503..783S} Stratta, G., Pozanenko, A., Atteia, J.-L., et al.\ 2009, \aap, 503, 783
\bibitem[Swenson, \& Roming(2014)]{2014ApJ...788...30S} Swenson, C.~A., \& Roming, P.~W.~A.\ 2014, \apj, 788, 30.
\bibitem[Troja et al.(2008)]{2008MNRAS.385L..10T} Troja, E., King, A.~R., O'Brien, P.~T., Lyons, N., \& Cusumano, G.\ 2008, \mnras, 385, L10 
\bibitem[Troja et al.(2016)]{2016ApJ...827..102T} Troja, E., Sakamoto, T., Cenko, S.~B., et al.\ 2016, \apj, 827, 102 
\bibitem[Troja et al.(2018)]{22652} Troja, E., Butler, N., Watson, A.~M., et al.\ 2018, GRB Coordinates Network, Circular Service, No.~22652, \#1 (2018/April-0), 22652, 1 
\bibitem[Troja et al.(2017)]{2017Natur.547..425T} Troja, E., Lipunov, V.~M., Mundell, C.~G., et al.\ 2017, \nat, 547, 425
\bibitem[Troja et al.(2018)]{22664} Troja, E., Butler, N., Watson, A.~M., et al.\ 2018, GRB Coordinates Network, Circular Service, No.~22664, \#1 (2018/April-0), 22664, 1
\bibitem[Troja et al.(2019)]{2019arXiv190501290T} Troja, E., Castro-Tirado, A.~J., Becerra Gonzalez, J., et al.\ 2019, arXiv:1905.01290 
\bibitem[van der Horst et al.(2014)]{2014MNRAS.444.3151V} van der Horst, A.~J., Paragi, Z., de Bruyn, A.~G., et al.\ 2014, \mnras, 444, 3151.
\bibitem[Vestrand et al.(2014)]{2014Sci...343...38V} Vestrand, W.~T., Wren, J.~A., Panaitescu, A., et al.\ 2014, Science, 343, 38
\bibitem[Vreeswijk et al.(2008)]{2008GCN..7444....1V} Vreeswijk, P.~M., Smette, A., Malesani, D., et al.\ 2008, GRB Coordinates Network, 7444, 1 
\bibitem[Watson et al.(2012)]{2012SPIE.8444E..5LW} Watson, A.~M., Richer, M.~G., Bloom, J.~S., et al.\ 2012, \procspie, 8444, 84445L 
\bibitem[Wei(2003)]{2003A&A...402L...9W} Wei, D.~M.\ 2003, \aap, 402, L9 
\bibitem[Wei et al.(2006)]{2006ApJ...636L..69W} Wei, D.~M., Yan, T., \& Fan, Y.~Z.\ 2006, \apjl, 636, L69 
\bibitem[Yasuda et al.(2001)]{2001AJ....122.1104Y} Yasuda, N., Fukugita, M., Narayanan, V.~K., et al.\ 2001, \aj, 122, 1104 
\bibitem[Yi et al.(2017)]{2017ApJ...844...79Y} Yi, S.-X., Yu, H., Wang, F.~Y., et al.\ 2017, \apj, 844, 79.
\bibitem[Yu et al.(2009)]{2009ApJ...692.1662Y} Yu, Y.~W., Wang, X.~Y., \& Dai, Z.~G.\ 2009, \apj, 692, 1662 
\bibitem[Zhang \& M{\'e}sz{\'a}ros(2002)]{2002ApJ...566..712Z} Zhang, B., \& M{\'e}sz{\'a}ros, P.\ 2002, \apj, 566, 712 
\bibitem[Zhang \& M{\'e}sz{\'a}ros(2004)]{2004IJMPA..19.2385Z} Zhang, B. and M{\'e}sz{\'a}ros, P.\ 2004, International Journal of Modern Physics A, 19, 2385
\bibitem[Zhang et al.(2006)]{2006ApJ...642..354Z} Zhang, B., Fan, Y.~Z., Dyks, J., et al.\ 2006, \apj, 642, 354 
\bibitem[Zhang et al.(2015)]{2015ApJ...798....3Z} Zhang, S., Jin, Z.-P., \& Wei, D.-M.\ 2015, \apj, 798, 3.
\bibitem[Zhang et al.(2018)]{Zhang18} Zhang, B.-B., Zhang, B., Castro-Tirado, A.~J., et al.\ 2018, Nature Astronomy, 2, 69 
\bibitem[Zheng \& Filippenko(2018)]{22647} Zheng, W. \& Filippenko, A.~V., 2018, GRB Coordinates Network, 22647, 1

\end{thebibliography}
\end{document}